\documentclass[journal=jacsat,manuscript=article]{achemso}
\setkeys{acs}{etalmode=truncate,maxauthors=10}
\usepackage[version=3]{mhchem} 
\usepackage{xcolor}
\usepackage{soul}

\usepackage{amsmath}
\usepackage{amssymb}
\usepackage{bm}           
\usepackage{mathrsfs}
\usepackage{booktabs,enumitem} 
\usepackage[footnotehyper]{nicematrix} 
\usepackage{multirow}
\usepackage{float}
\usepackage{etoolbox}  

\usepackage[labelfont=bf, skip=0pt]{caption} 

\usepackage{comment}

\makeatletter
\patchcmd{\acs@contact@details}{E-mail}{*\,Email}{}{}  
\makeatother

\author{Justin Airas}
\author{Bin Zhang}
\email{binz@mit.edu}
\affiliation{Department of Chemistry, Massachusetts Institute of Technology, Cambridge, MA, USA}

\title[An \textsf{achemso} demo]
  {Scaling Graph Neural Networks to Large Proteins}

\begin{document}

\begin{abstract}
Graph neural network (GNN) architectures have emerged as promising force field models, exhibiting high accuracy in predicting complex energies and forces based on atomic identities and Cartesian coordinates. To expand the applicability of GNNs, and machine learning force fields more broadly, optimizing their computational efficiency is critical, especially for large biomolecular systems in classical molecular dynamics simulations. In this study, we address key challenges in existing GNN benchmarks by introducing a dataset, DISPEF, which comprises large, biologically relevant proteins. DISPEF includes 207,454 proteins with sizes up to 12,499 atoms and features diverse chemical environments, spanning folded and disordered regions. The implicit solvation free energies, used as training targets, represent a particularly challenging case due to their many-body nature, providing a stringent test for evaluating the expressiveness of machine learning models. We benchmark the performance of seven GNNs on DISPEF, emphasizing the importance of directly accounting for long-range interactions to enhance model transferability. Additionally, we present a novel multiscale architecture, termed Schake, which delivers transferable and computationally efficient energy and force predictions for large proteins. Our findings offer valuable insights and tools for advancing GNNs in protein modeling applications.
\end{abstract}

\clearpage
\newpage
\section{Synopsis}
We introduce the DISPEF dataset to benchmark the performance of graph neural networks on large proteins and propose a novel multiscale architecture that enables efficient and precise protein modeling.

\section{Introduction}
    Machine learning (ML) force fields have garnered considerable attention in recent years due to their ability to account for many-body effects with higher accuracy and computational efficiency compared to \emph{ab initio} methods.\cite{airas_transferable_2023,majewski_machine_2023,katzberger_implicit_2023,katzberger_general_2024,chen_machine_2021,husic_coarse_2020,kozinsky_scaling_2023,wang_improving_2023,brunken_machine_2024,yao_machine_2023,wang_machine_2019,wang_multi-body_2021,fu_forces_2023,durumeric_learning_2024,arts_two_2023,kohler_flow-matching_2023,loose_coarse-graining_2023,durumeric_adversarial-residual-coarse-graining_2019,durumeric_machine_2023,ding_contrastive_2022,ding_optimizing_2024,chmiela_towards_2018,chmiela_accurate_2020,duschatko_thermodynamically_2024,ggreener_differentiable_2024,yang_slicing_2023,corso_graph_2024,batzner_advancing_2023,eastman_openmm_2024,zheng_predicting_2024,sahrmann_utilizing_2023,bonneau_peering_2024,anstine_machine_2023,cheng_developing_2024,faller_density-based_2024,takaba_machine-learned_2024,wang_espalomacharge_2024,wang_end--end_2022,galvelis_nnpmm_2023,sabanes_zariquiey_enhancing_2024,barnett_neural_2024} Given that molecules are naturally structured as graphs, with atoms represented as nodes and interactions as edges, graph neural networks (GNNs) offer a compelling framework for force field modeling. These networks also facilitate the straightforward incorporation of physical symmetries, such as translational, rotational, and permutational symmetries, further boosting their appeal.

    Although numerous GNN architectures have been proposed,\cite{unke_physnet_2019,gasteiger_directional_2022,gasteiger_gemnet_2022,anderson_cormorant_2019,brandstetter_geometric_2022,li_egnn_2022,schutt_equivariant_2021,thomas_tensor_2018,fuchs_se3-transformers_2020,huang_equivariant_2022,schutt_schnet_2018,satorras_en_2022,wang_spatial_2023,tholke_torchmd-net_2022,pelaez_torchmd-net_2024,wang_enhancing_2024,batzner_e3-equivariant_2022,batatia_design_2022,musaelian_learning_2023,batatia_mace_2023} they generally follow a similar operational principle. Initially, feature vectors are assigned to atoms based on their type. These atomic features are then updated via the message-passing layers, which incorporate information from neighboring atoms. Specifically, each message-passing layer, totaling $N_\mathrm{layers}$, updates atomic features by aggregating data from neighboring atoms $j$ within a cutoff radius $r_\mathrm{cut}$ from atom $i$. As this process repeats across layers, increasingly distant atomic information is incorporated. Finally, the updated atomic features are transformed into contributions to the total potential energy, while force predictions are obtained by calculating the negative gradient of the predicted energy with respect to the atomic positions.

    The development of GNN architectures has traditionally prioritized accuracy and expressiveness, with limited attention to computational efficiency. Although ML-based force fields are generally more efficient than quantum mechanical methods, they remain significantly slower compared to highly optimized classical force fields. Notably, several studies have applied GNNs to protein molecules in MD simulations\cite{airas_transferable_2023,majewski_machine_2023,katzberger_implicit_2023,katzberger_general_2024,chen_machine_2021,husic_coarse_2020,kozinsky_scaling_2023,wang_improving_2023,brunken_machine_2024,yao_machine_2023,wang_machine_2019,wang_multi-body_2021,fu_forces_2023,durumeric_learning_2024,arts_two_2023,wang_enhancing_2024}, demonstrating impressive accuracy but encountering limited practical utility due to high computational costs. Consequently, it is essential to systematically assess the performance of GNN architectures on large biomolecules to identify optimal designs and potentially guide future advancements.

    Benchmarking the performance of GNN architectures on large biomolecules is complicated by the lack of datasets.
    Most existing datasets such as QM9\cite{ramakrishnan_quantum_2014}, 
    ISO17\cite{schutt_schnet_2018}, 
    MD17\cite{chmiela_machine_2017}, 
    the ANI datasets\cite{smith_ani-1_2017,smith_ani-1ccx_2020}, 
    the SPICE datasets\cite{eastman_spice_2023,eastman_nutmeg_2024},
    and the ZINC dataset\cite{sterling_zinc_2015}, contain only small molecules. 
    While datasets like COMP6v2\cite{smith_less_2018,devereux_extending_2020} and MD22\cite{chmiela_accurate_2023} contain a handful of larger structures, they are still relatively limited in their representation of large, biologically-relevant molecules.
    Due to these dataset limitations, previous GNN benchmarking studies\cite{wu_moleculenet_2018,sorkun_aqsoldb_2019,modee_benchmark_2022,dwivedi_benchmarking_2022,fu_forces_2023} have not focused on large biomolecules.

    In this study, we address the challenges of benchmarking and developing GNN architectures for large, biologically relevant proteins. We first present a dataset, DISPEF (\textbf{D}ataset of \textbf{I}mplicit \textbf{S}olvation \textbf{P}rotein \textbf{E}nergies and \textbf{F}orces), which contains implicit solvation free energies for proteins ranging in size from 16 to 1,022 amino acids. The dataset includes both folded and disordered regions, offering a diverse array of chemical environments. Along with the many-body nature of solvation free energies, these environments present challenging cases for evaluating the effectiveness of ML approaches in mapping atomistic representations to energy. We then benchmark several recent GNN architectures using DISPEF, providing critical insights into key design features essential for ensuring model accuracy and transferability. Building on these benchmark results, we introduce a hybrid, multiscale GNN architecture that demonstrates high efficiency and transferability, particularly to proteins significantly larger than those in the training set. 
    This custom GNN architecture outperforms all existing models examined in this work. Overall, our study offers valuable tools and insights to facilitate the advancement of GNN architectures tailored for protein applications.

\section{Results}

\subsection{DISPEF for benchmarking GNN performance on large proteins}

    Much of the recent effort to develop more accurate GNNs has largely focused on minimizing prediction errors on widely-used datasets of small, drug-like molecules.
    However, to develop GNNs that are capable of efficiently and transferably scaling to large and complex proteins containing thousands of atoms, datasets of large protein structures are necessary. 
    To address this data limitation, we've constructed a dataset of implicit solvation protein energies and forces (abbreviated as DISPEF) that includes over 200,000 proteins ranging in size from 16 to 1,022 amino acids. 
    This dataset contains AlphaFold2\cite{jumper_highly_2021,varadi_alphafold_2022}-predicted structures for proteins within the Swiss-Prot database.\cite{the_uniprot_consortium_uniprot_2023} 
    Since the Swiss-Prot database is hand-curated to minimize sequence redundancy, and since structure predictions from AlphaFold2 contain both higher-confidence folded regions and lower-confidence unfolded / disordered regions, DISPEF covers a wide range of chemical environments.

\begin{figure}[t!]
    \centering
    \includegraphics[width=\textwidth]{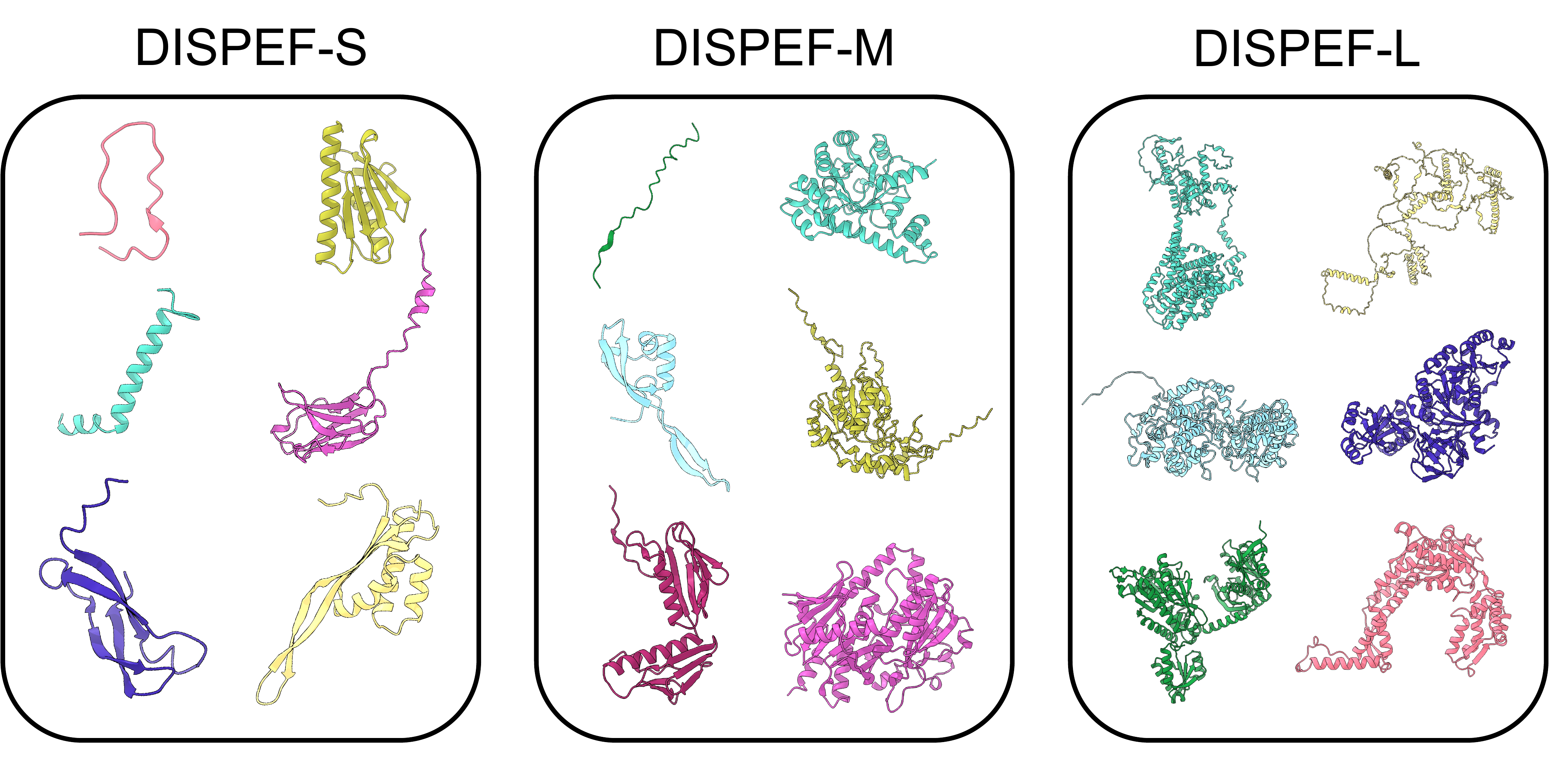}
    \caption{
    \textbf{DISPEF and its subsets were constructed to evaluate GNN architectures.}
    The first subset (left, denoted DISPEF-S) contains $\sim$ 81,000 proteins with under 2,000 atoms. The second subset (center, denoted DISPEF-M) contains 24,000 proteins with under 400 amino acids. The third subset (right, denoted DISPEF-L) contains over 109,000 proteins ranging in size from 6,800 to 12,499 atoms. A smaller fourth subset, DISPEF-c (not pictured), was constructed to evaluate the computational cost of GNNs.
    }
    \label{dataset_figs}
\end{figure}

    In addition to all-atom protein structures, we incorporate solvation free energies and the corresponding atomic forces, computed using the GBn2 model \cite{mongan_generalized_2007,nguyen_improved_2013}, as target values in DISPEF. Although existing databases often provide quantum mechanical energies and forces \cite{wang_aimd-chig_2023,wang_enhancing_2024}, calculating these for proteins of biologically relevant size is computationally prohibitive. However, unlike many of the pairwise energy terms commonly used in MD simulations, GB-based implicit solvent models utilize many-body potentials to estimate solvation free energy, accounting for both electrostatic and non-polar contributions \cite{onufriev_generalized_2019}. The long-range and many-body characteristics of these models offer robust benchmarks for assessing the capability of GNNs to learn complex relationships that map atomic structures to energy.

    We created four subsets of DISPEF that aim to assess different capabilities of GNNs. Representative structures from three of these subsets are displayed in Figure \ref{dataset_figs}. Throughout this manuscript, we'll refer to the subsets as follows:
    \begin{enumerate}
        \item DISPEF-S - this consists of 81,341 proteins with less than 2,000 atoms.
        This subset was further split into training and testing sets consisting of 65,072 and 16,269 proteins, respectively. 
        \item DISPEF-M - this consists of 24,000 proteins with less than 400 amino acids ($\sim$ 6,800 atoms). 
        This subset was further split into training and testing sets consisting of 19,200 and 4,800 proteins, respectively. 
        \item DISPEF-L - this consists of 109,108 proteins with greater than 6,800 atoms and less than 12,499 atoms. 
        \item DISPEF-c - this consists of 560 randomly selected proteins, spaced roughly evenly in size from 192 to 12,346 atoms. 
    \end{enumerate}
    DISPEF-S aims to assess the ability of GNNs to produce accurate predictions across a large quantity of relatively small proteins, while DISPEF-M aims to assess the scalability of GNNs across a smaller set of different sized proteins. DISPEF-L aims to assess the transferability of GNNs to structures significantly larger than those in the training set. While DISPEF-L could also be used to train GNNs, this would likely require a GPU with greater than 32 GB of memory. Lastly, the smaller DISPEF-c was constructed to evaluate the computational cost of GNNs. 
    Overall, the construction of DISPEF enables comprehensive investigation into the accuracy, scalability, and transferability of GNN architectures to large, biologically-relevant proteins.

\subsection{Selection of diverse and efficient GNN architectures}

    With the creation of DISPEF, we can now benchmark the prediction accuracy of various GNN architectures on large, biologically-relevant proteins. We focus on a wide collection of recently introduced GNNs with diverse architectural designs, whose superior performances and computational efficiency have been documented in prior studies. 
    As such, the SchNet\cite{schutt_schnet_2018}, Graph Network (GN)\cite{majewski_machine_2023}, EGNN\cite{satorras_en_2022}, SAKE\cite{wang_spatial_2023}, Equivariant Transformer (ET)\cite{tholke_torchmd-net_2022,pelaez_torchmd-net_2024},  and ViSNet\cite{wang_enhancing_2024} architectures were selected. While GNN architectures that utilize Clebsh-Gordon tensor products to achieve equivariance have proven to be highly accurate for small molecules\cite{batzner_e3-equivariant_2022,batatia_design_2022,musaelian_learning_2023,batatia_mace_2023}, they were not investigated in this study due to their well-attested computational cost \cite{wang_enhancing_2024}. 
    Additionally, while not a GNN, we also included ANI\cite{smith_ani-1_2017,smith_ani-1ccx_2020,devereux_extending_2020} in this study due to its computational efficiency and its similar use-cases to GNNs.

    While we point to the work of \citet{han_survey_2024} for a comprehensive overview of GNN architectures, we'll briefly detail some of the differences between the selected GNNs. 
    The SchNet architecture utilizes continuous filter convolutions acting on radial basis expansions of pairwise distances $r_{ji}$ between atom $i$ and its neighboring atoms $j$ to compute the feature update for atom $i$.\cite{schutt_schnet_2018} 
    The GN architecture is a modified SchNet architecture that incorporates a radial basis function with trainable parameters and an initial atomic featurization that incorporates information from neighboring atoms.\cite{majewski_machine_2023} 
    By operating only on the distances between atoms, both SchNet and GN predictions are invariant with respect to $E(3)$ transformations, corresponding to translational, rotational, and reflectional symmetry.

    Recent GNN developments have focused on equivariant designs, where predictions will undergo equivalent transformations when a certain transformation was performed on the input. Such designs can be useful for predicting vectorial quantities such as forces or dipole moments, whose directions depend on the molecule orientation. 
    The EGNN architecture utilizes edge features defined between atoms $i$ and $j$ to compute the feature update for atom $i$, and can additionally be made equivariant by performing a coordinate transformation in each message-passing layer.\cite{satorras_en_2022} 
    The SAKE architecture is a recent improvement to EGNN that incorporates attention acting on edge features along with SchNet-style continuous filter convolutions to compute the feature update for atom $i$.\cite{wang_improving_2023} 
    We opted not to perform coordinate transformation within EGNN and SAKE, as this made training unstable for the large proteins that we focused on in this study. As such, we used $E(3)$-invariant versions of EGNN and SAKE.
    The ET architecture is an equivariant graph transformer that utilizes atomic features and pairwise distances $r_{ji}$ to compute the query, key, and value needed to compute attention weights and then the feature update for atom $i$.\cite{tholke_torchmd-net_2022} ViSNet, the most recent of the architectures we selected, is an equivariant GNN that utilizes dot products between pairwise displacement unit vectors and between vector rejections to incorporate angular and torsional information.\cite{wang_enhancing_2024}

    Lastly, ANI computes feature vectors using the atom types, radial, and angular information from neighboring atoms, and then transforms those feature vectors into energy contributions using an ensemble of feed-forward neural networks (one per atom type).\cite{smith_ani-1_2017,gao_torchani_2020,devereux_extending_2020}

\subsection{Larger cutoff distances are required for transferable predictions}
    We next evaluated the energy prediction accuracy of 6 different GNNs and ANI using our subsets of DISPEF. Performance of GNNs can be highly dependent on hyperparameters, so we carried out a systematic hyperparameter optimization for each architecture. Due to the computational cost for training on DISPEF, the optimization was separated into two stages. First, we optimized on a dataset of MD configurations and their DFT-level energies and forces for chignolin CLN025.\cite{honda_crystal_2008,wang_aimd-chig_2023} Methodological details of this training are discussed in the Supporting Information, and prediction errors can be viewed in Table S1.
    Then we optimized over DISPEF-S. To reduce training time as much as possible, we implemented a custom, variable-size batch sampler to ensure optimal VRAM usage during training (detailed in the Methods section). From hyperparameter optimization, we found that the number of message-passing layers ($N_\mathrm{layers}$) and the cutoff distance for identifying neighboring atoms ($r_\mathrm{cut}$) had the most impact on model accuracy and computational efficiency. Both hyperparameters are common to all GNNs investigated in this study. All explored and optimal hyperparameters for all models can be viewed in Tables S2 through S9.

\begin{figure}[t]
    \centering
    \includegraphics[width=0.9\textwidth]{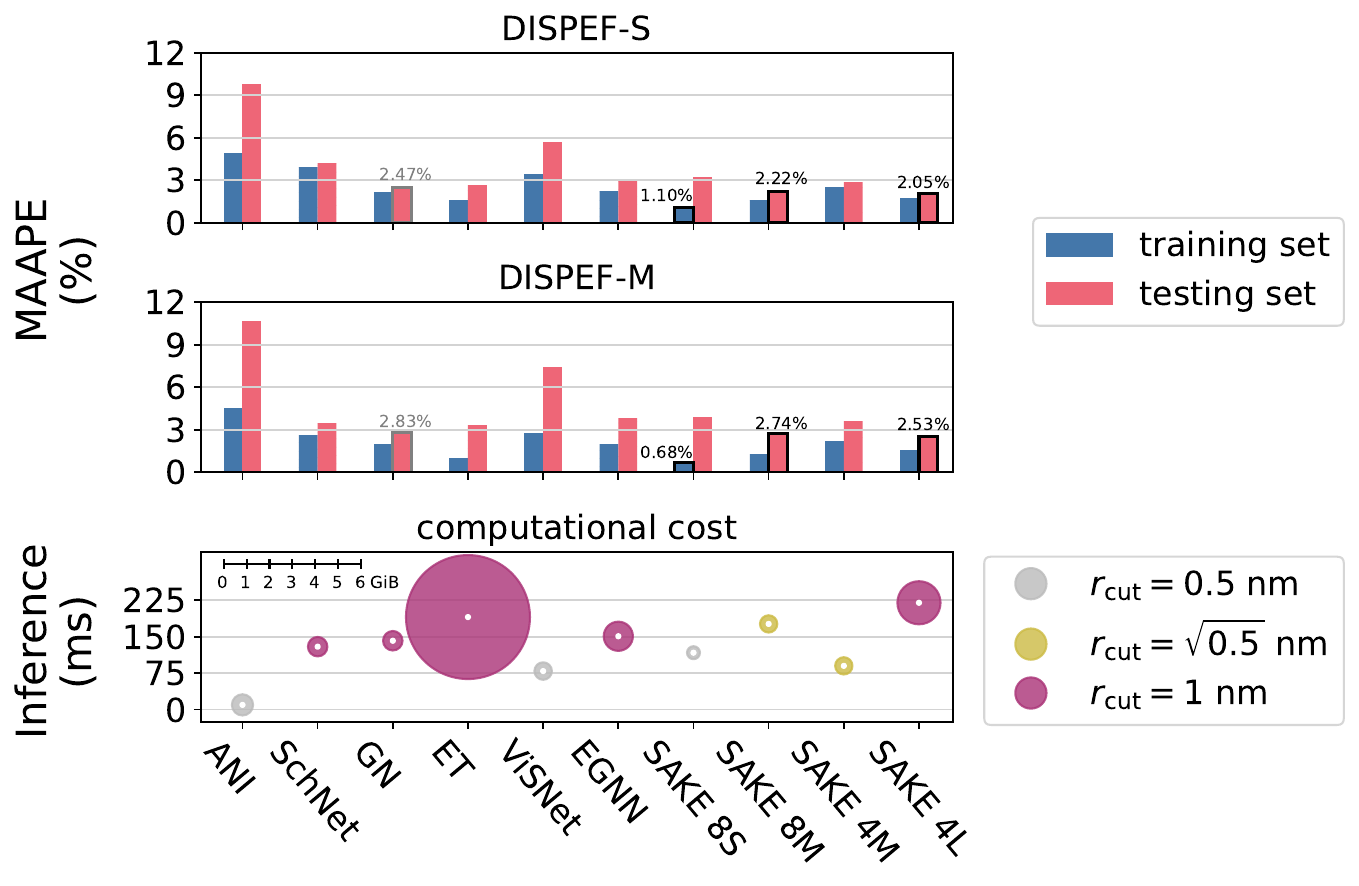}
    \caption{
    \textbf{
    Benchmarking GNN performance on large proteins highlights the advantages of SAKE architectural innovation and significance of increasing $\bm{r}_\text{cut}$ for transferability.
    }
    In the top two figures, we show the energy mean arctangent absolute percent error\cite{kim_new_2016} (MAAPE) for all models on DISPEF-S and DISPEF-M.
    The most accurate models are outlined in black, while the testing set errors for GN (which had achieved the lowest testing set errors before we performed further hyperparameter optimization for SAKE) are outlined in gray. 
    MAAPEs for all discussed models are displayed on the figure. Models with $r_\mathrm{cut} = 0.5$ nm are depicted with solid bars, while models with $r_\mathrm{cut} = \sqrt{0.5}$ and 1 nm are depicted with increasingly more dense diagonal lines through the bars. In the bottom figure, inference times and peak memory allocation as measured on the representative structure of DISPEF-M (UniProt ID: A1W0R3) are displayed. Peak memory allocation is proportional to the width of each circle.
    }
    \label{E_swissprot}
\end{figure}

    With the optimal hyperparameters, we report the accuracy of each GNN on training and testing sets in the top two panels of Figure \ref{E_swissprot}.
    Since the energy value inherently increases with increasing protein size, 
    we report the 
    mean arctangent absolute percent error (MAAPE)\cite{kim_new_2016} as a relative error metric (detailed in the Supporting Information). 
    The mean absolute error (MAE) and the $r^2$ value for these models are displayed in Tables S10 and S11. 
    Additionally, we used DISPEF-c to conduct a detailed analysis of the average inference times and maximum VRAM allocations of each model with respect to protein size. Discussion of this analysis can be found in the Supporting Information, along with Figures S3 through S6. For simplicity, however, we show the average inference time and maximum VRAM allocation for a representative structure from DISPEF-M in the bottom panel of Figure \ref{E_swissprot}.

    Our initial results suggested that the GN architecture 
    performed best on both the DISPEF-S and DISPEF-M testing sets, while also being relatively computationally efficient (inference time: 141.73 ms, VRAM cost: 0.94 GiB). Notably, the ViSNet architecture, which has demonstrated state-of-the-art energy and force prediction for small molecules and CLN025,\cite{wang_enhancing_2024} showed the weakest performance among the GNNs evaluated in this study. Although further hyperparameter optimization improved accuracy, ViSNet continued to perform suboptimally (Table S3). Enhancing ViSNet's accuracy would likely require increasing the spherical harmonics degree, which is impractical due to the substantial VRAM requirements.

    Our first hyperparameter combination for SAKE ($N_\mathrm{layers} = 8$, $r_\mathrm{cut} = 0.5$ nm), referred to as SAKE 8S, achieved the lowest error on the training sets. However, it exhibited a notable decline in accuracy on the test sets, with MAAPE decreases of 2.13\% and 3.22\% for DISPEF-S and DISPEF-M, respectively. Other architectures (ANI, ET, and ViSNet) also showed large accuracy declines. This performance drop, particularly on the larger proteins in DISPEF-M, suggests that these models may have overfit the training data. These neural networks appear to approximate long-range interaction energies using local atomic environments within the cutoff distance, rather than accurately assigning them to atoms that are spatially distant. This approximation of long-range interactions results in decreased accuracy on the testing set.

    The computational efficiency of SAKE together with its training set accuracy made it a promising candidate for further hyperparamter optimization. As such, further combinations of $N_\mathrm{layers}$ and $r_\mathrm{cut}$ were explored.
    An increase in $r_\mathrm{cut}$ to $\sqrt{0.5} \approx 0.707$ nm produced a SAKE model (referred to as SAKE 8M) that surpassed GN in accuracy on both the DISPEF-S and DISPEF-M test sets. However, this came at the cost of higher computational demands, with an inference time of 176.38 ms and a VRAM usage of 0.83 GiB. Reducing $N_\mathrm{layers}$ from 8 to 4, while maintaining $r_\mathrm{cut} = \sqrt{0.5}$ nm, led to a more computationally efficient model (SAKE 4M), reducing inference time to 90.01 ms and preserving the same VRAM usage, though with a trade-off in accuracy. Further increasing $r_\mathrm{cut}$ to 1 nm yielded a model (SAKE 4L) with the best performance on the test sets. 
    Nonetheless, the increased computational cost of SAKE 4L, with an inference time of 219.92 ms and a VRAM requirement of 2.07 GiB, emphasizes that while incorporating longer-range information can enhance accuracy, simply increasing $r_\mathrm{cut}$ is not a computationally scalable solution.

\subsection{A multiscale GNN architecture enables transferable and efficient predictions}

    Our benchmark results suggest that efficient and transferable GNNs need to operate at large cutoff distances with low computational cost. We propose a multiscale architectural design that meets these two criteria. 
    Specifically, we make use of 
    two different message-passing layers that operate on different graphs / subgraphs.
    A highly accurate SAKE layer updates each atomic featurization using information from all short-ranged neighbors determined by cutoff distance $r_\mathrm{cut}^\mathrm{SAKE}$, followed by a less accurate but more efficient SchNet layer to compute a second update to each atomic featurization using information from long-ranged neighboring alpha carbon atoms that fall in the range of $r_\mathrm{cut}^\mathrm{SchNet}$. 
    We call this mixed model Schake (pronounced as ``shock'' followed by the letter ``A'', or ``shock-A''), since it uses both SchNet and SAKE message-passing layers. 
    Recently, a similar design was used in Chroma\cite{ingraham_illuminating_2023}, a denoising probabilistic diffusion model for generating atomistic protein conformations.

\begin{figure}[t]
    \centering
    \includegraphics[width=0.8\textwidth]{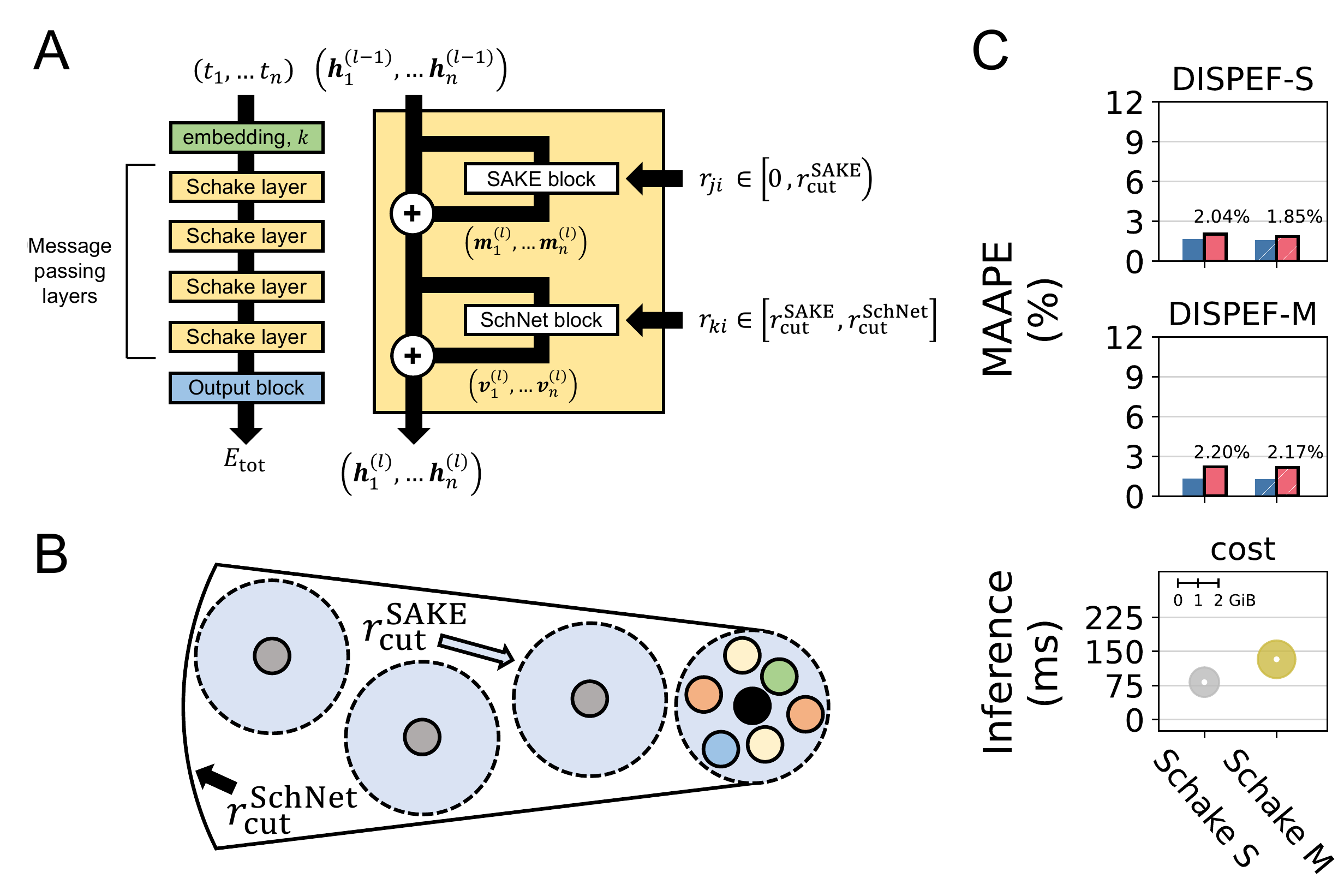}
    \caption{
    \textbf{The Schake architecture enables accurate and efficient incorporation of longer-ranged interactions.}
    (A) Illustration of the Schake architecture. Architectural details are explained in the Methods section and the Supporting Information.
    (B) Illustration of the multiscale resolution of Schake. Dashed blue space represents the short-ranged environments processed by the SAKE layer, while the white space represents the long-ranged environment processed by the SchNet layer. The solid black circle represents atom $i$, the solid multi-colored circles represent the many atom types of short-ranged neighbors $j$ processed by the SAKE layer, and the solid gray circles represent the alpha carbon atom types of long-ranged neighbors $k$ processed by the SchNet layer.
    (C) Energy MAAPE results for DISPEF-S and DISPEF-M along with inference and peak memory allocation measurements for the representative structure of DISPEF-M (UniProt ID: A1W0R3). These figures are formatted identically to Figure \ref{E_swissprot}.
    }
    \label{Schake_diagram}
\end{figure}

    The multiscale design significantly expands the attention range of each atom, allowing them to access chemical information at a much longer range. 
    Since short-ranged neighbors are still used to produce their feature updates, each alpha carbon atom in the sparse subgraph would still contain information from its surrounding chemical environment (Figure \ref{Schake_diagram}B). Collecting information from alpha carbons allow the atom under consideration to experience the atomic environment at a distance set by $r_\mathrm{cut}^\mathrm{SchNet}$, rather than the much smaller value $r_\mathrm{cut}^\mathrm{SAKE}$.

    In addition, since a majority of the atom pairs will be processed by the more computationally efficient architecture, SchNet, Schake maintains efficiency. 
    As shown by Figure S1A, for the Schake model with $r_\mathrm{cut}^\mathrm{SAKE} = 0.5$ nm and $r_\mathrm{cut}^\mathrm{SchNet} = 2.5$ nm,
    the SAKE message-passing layer processes $\sim$ 20 - 25\% of the total atom pairs, while the SchNet layers processes the remaining $\sim$ 75 - 80\% of atom pairs
    for structures larger than $\sim$ 3,000 atoms. Furthermore, the total number of processed atom pairs in Schake remains smaller 
    than any model with $r_\mathrm{cut} = 1$ nm (Figure S1B), despite a cutoff distance of 2.5 nm.

    We evaluated the performance of two Schake models, with $r_\mathrm{cut}^\mathrm{SAKE} = 0.5$ or $\sqrt{0.5}$ nm, and $r_\mathrm{cut}^\mathrm{SchNet} = 2.5$ nm (denoted as Schake S and Schake M, respectively). As shown in Figure \ref{Schake_diagram}C, both Schake models are more accurate than all other models across the two testing sets. This accuracy improvement is coupled with a significant decrease in inference times compared to the previously best performing models. 
    Specifically, the Schake models are 2.7$\times$ (82.25 ms) and 1.7$\times$ (132.50 ms) faster than the previous most accurate model, SAKE 4L (219.92 ms). 
    While the peak memory allocation of Schake is still relatively high, we discuss causes for this in the Supporting Information. Overall, these results suggest that the Schake architecture is both highly accurate and highly computationally efficient.

\begin{figure}[t]
    \centering
    \includegraphics[width=0.9\textwidth]{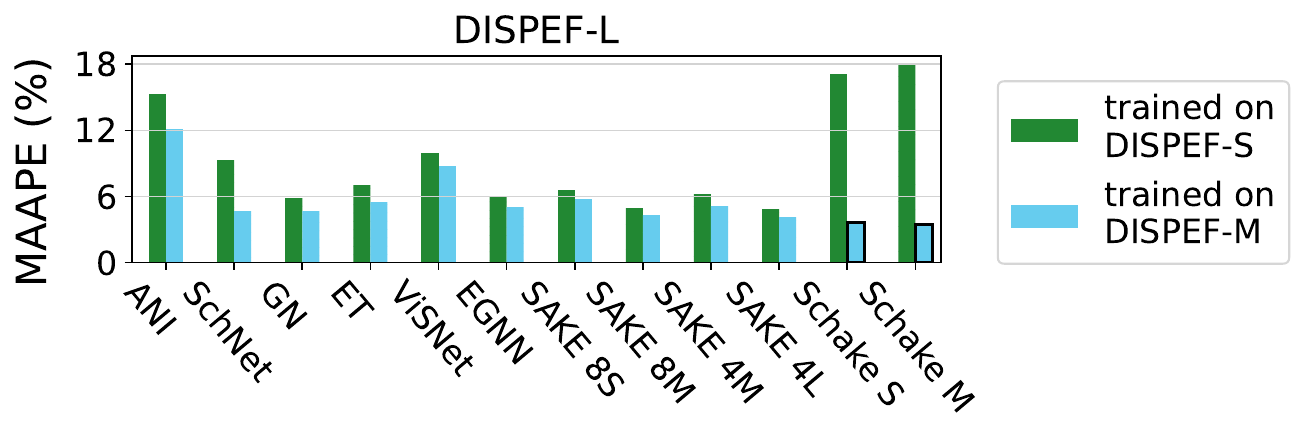}
    \caption{
    \textbf{Training on DISPEF-M results in models that are more transferable to the larger proteins of DISPEF-L.}
    Models with $r_\mathrm{cut} = 0.5$ nm are depicted with solid bars, while models with $r_\mathrm{cut} = \sqrt{0.5}$ and 1 nm are depicted with increasingly more dense diagonal lines through the bars.
    For all models, we show the mean arctangent absolute percent error\cite{kim_new_2016} (MAAPE). The best-performing models are outlined in black. While the Schake model trained on DISPEF-S performs poorly on DISPEF-L, the Schake model trained on DISPEF-M achieves the lowest errors of all tested models on DISPEF-L.
    }
    \label{val_preds}
\end{figure}

\subsection{Schake enables transferable and scalable energy and force predictions}

    For ML models to be useful, they should ideally be transferable to both similar-sized and larger proteins outside of the training set. 
    To this end, we next assessed the transferability of Schake and other models to the larger proteins in DISPEF-L. As detailed earlier, DISPEF-L contains over 109,000 proteins ranging in size from 6,800 to 12,499 atoms. Using the two models trained on either DISPEF-S or DISPEF-M for each architecture, we computed energy predictions for all proteins within DISPEF-L. Prediction errors can be viewed in Figure \ref{val_preds}. These results suggest that across all architectures, greater accuracy for the larger proteins of DISPEF-L is achieved when the training set contains larger structures (i.e., when trained on DISPEF-M). This is especially pronounced with the Schake models, where the larger cutoff distance of $r_\mathrm{cut}^\mathrm{SchNet} = 2.5$ nm is not fully utilized when training on DISPEF-S. 
    Across the models trained on DISPEF-S, SAKE 4L achieves the best accuracy on DISPEF-L. However, across the models trained on DISPEF-M, both Schake models achieve the best accuracy on DISPEF-L.

\begin{figure}[t]
    \centering
    \includegraphics[width=0.95\textwidth]{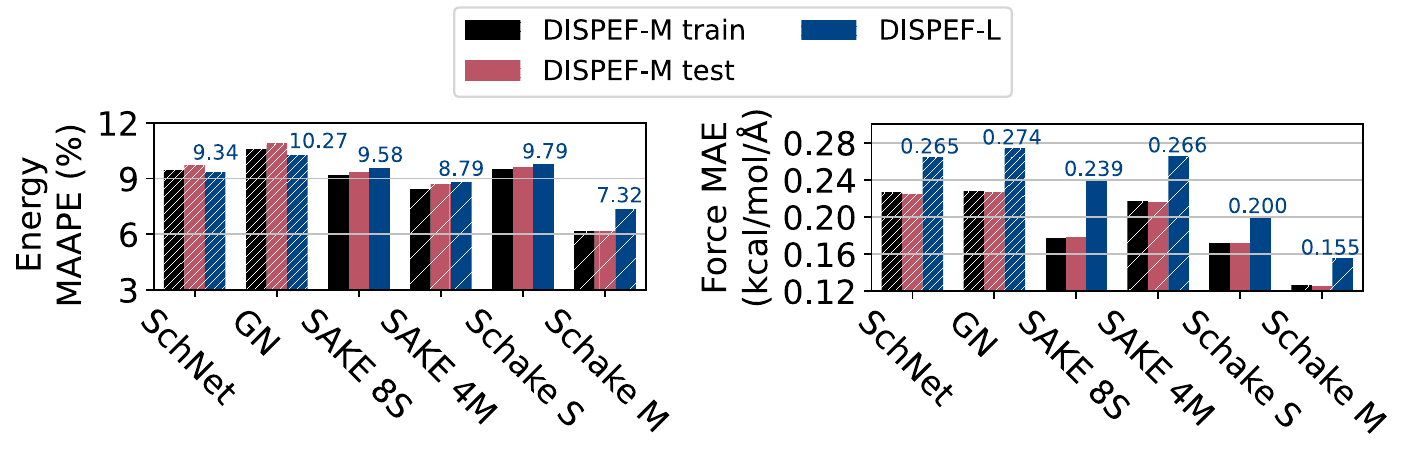}
    \caption{
    \textbf{The Schake models outperform SchNet, GN, and SAKE at force prediction.}
    (Left) Energy prediction mean arctangent absolute percent error\cite{kim_new_2016} (MAAPE). (Right)
    Force mean absolute error (MAE). All DISPEF-L energy and force errors are shown in blue text.
    Models with $r_\mathrm{cut} = 0.5$ nm are depicted with solid bars, while models with $r_\mathrm{cut} = \sqrt{0.5}$ and 1 nm are depicted with increasingly more dense diagonal lines through the bars.
    }
    \label{force_errors}
\end{figure}

    When using a GNN as an MD force field, force prediction accuracy is especially important to ensure accurate dynamics. To assess force prediction accuracy, we trained a selection of the most computationally efficient of the architectures we explored on both the energies and forces of structures in DISPEF-M. Due to memory limitations, we could not perform this training on the closest performing model to Schake, SAKE 4L.
    Energy and force training for each model was conducted using the same hyperparameters used for the energy-only training reported above.
    Prediction errors from this training are shown in Figure \ref{force_errors}. 
    Note that energy errors contribute little to the loss function used during force and energy training (Figure S2). As such, it is not expected for these energy errors to match the results shown in Figures \ref{E_swissprot} through \ref{val_preds}.
    Despite this, Schake M achieves the lowest energy errors of the tested models. Schake S achieves similar energy errors to SAKE 8S. However, both Schake models outperform all of the other GNNs explored in this study in force prediction accuracy. Schake S is more accurate than the next most accurate model, SAKE 8S, while also being $\sim$ 1.4$\times$ faster. Additionally, Schake M is substantially more accurate than SAKE 8S while only being $\sim$ 1.1$\times$ slower. These results again demonstrate the success of our Schake architecture.

\section{Conclusions}
    We have presented DISPEF and its four subsets, named DISPEF-S, DISPEF-M, DISPEF-L, and DISPEF-c, that enable thorough evaluation of the performance of GNN architectures for proteins. Energy inference results for the DISPEF subsets suggest that GNN architectures that are known to perform exceptionally well for small molecules are not necessarily guaranteed to also perform well for large proteins (Tables \ref{E_swissprot} and \ref{val_preds}). As such, DISPEF will help to guide future GNN architecture design toward more accurate predictions for large structures like proteins. Our results in Figure \ref{val_preds} show that longer-range interactions must be accounted for (in a computationally efficient manner) in order to achieve the most accurate predictions for proteins. To this end, we've also introduced a multiscale architecture termed ``Schake'' (Figure \ref{Schake_diagram}) that uses a more accurate but more expensive SAKE message-passing layer to process short-ranged neighbors, and a less accurate but more efficient SchNet message-passing layer to process longer-ranged alpha carbon neighbors.
    This mixed architecture design is not limited to just SAKE and SchNet; in the future, the SAKE layer could be replaced with an even-more accurate message-passing layer. Likewise, the SchNet layer could be replaced with a more efficient scheme to incorporate longer-ranged information.
    Our training and inference results show that Schake achieves state-of-the-art energy prediction on all of our DISPEF subsets (Figure \ref{Schake_diagram}C), and also achieves the smallest force errors when conducting energy and force training on DISPEF-M (Figure \ref{force_errors}).
    Overall, our results provide meaningful tools and insight for the future development of GNNs as broadly-applicable and scalable protein force fields and implicit solvent models.

\section{Methods}

\subsection*{Structures, energies, and forces for DISPEF}
    
    PDB structures for proteins in the Swiss-Prot database\cite{the_uniprot_consortium_uniprot_2023} were obtained from the AlphaFold2 protein structure database\cite{jumper_highly_2021,varadi_alphafold_2022}. 
    We next used OpenMM\cite{eastman_openmm_2017} (ver$.$ 7.7.0) to add hydrogen atoms to all structures.
    Energy minimization was then performed using OpenMM on the hydrogenated structures with the CHARMM36\cite{best_optimization_2012} force field and the GBn2 implicit solvent model\cite{nguyen_improved_2013}. For each minimized structure, we saved the positions ($\bm{x}$), the solvation free energy computed by the GBn2 implicit solvent ($E_\mathrm{GBn2}$), and the corresponding forces ($\bm{f}_\mathrm{GBn2} = -\nabla_{\bm{x}} E_\mathrm{GBn2}$). Units of nm were used for $\bm{x}$, units of kJ/mol were used for $E_\mathrm{GBn2}$, and units of kJ/mol/nm were used for $\bm{f}_\mathrm{GBn2}$. To ensure that the saved forces were exactly reproducible, the deterministic forces option for OpenMM's CUDA platform was set to true.
    We also saved the PDB atom names (for example, the label ``CA'' for alpha carbons) and the nuclear charges for all atoms in each structure. To enable scaling of the energy loss function by protein size, we also saved the total number of atoms in each structure ($n_\mathrm{atoms}$). 
    This data comprises DISPEF and its subsets. For all of these subsets, $\bm{x}$, $E_\mathrm{GBn2}$, $\bm{f}_\mathrm{GBn2}$, nuclear charges, PDB atom names, $n_\mathrm{atoms}$, and UniProt entry names were saved.
    For DISPEF-S, DISPEF-M, and DISPEF-L, $E_\mathrm{GBn2} >= -21,000$, $-21,000$, and $-25,000$ kcal/mol (respectively) for all structures.
    All subsets (saved as PyTorch dataset objects) and the code required to load them will be made publicly available (see Data Availability section).

\subsection*{Implementation of GNN Architectures}
    
    All models were implemented using PyTorch\cite{paszke_pytorch_2019} (ver$.$ 2.0) and PyTorch-Geometric\cite{fey_fast_2019} (ver$.$ 2.3.1). 
    We implemented ET\cite{tholke_torchmd-net_2022}, GN\cite{majewski_machine_2023}, and ViSNet\cite{wang_enhancing_2024} using the code provided by their respective publications / python packages. The torchANI\cite{gao_torchani_2020} implementation of ANI\cite{smith_ani-1_2017,smith_ani-1ccx_2020,devereux_extending_2020} was used, along with the CUDA-accelerated atomic environment vector computer. The ANI model used here consists of an ensemble of five 4-layer feed-forward NNs, corresponding to the five elements present in our dataset (H, C, N, O, and S).
    
    Our implementation of the SchNet architecture was based on that from the PyTorch-Geometric package. Minor changes inspired by the GN architecture were made to improve accuracy. The expnorm kernel function and the CELU (with $\alpha=2$) activation function were used, as opposed to the original Gaussian kernel function and the shifted softplus activation function. Our implementation of EGNN was based on the original PyTorch implementation. We modified this code to implement a cutoff distance ($r_\mathrm{cut}$).
    Since a PyTorch implementation of SAKE was not previously available, we created our own. We built the SAKE operations onto the EGNN PyTorch code.
    We further detail our implementation of the SAKE architecture in the Supporting Information.
    For both EGNN and SAKE, no velocity information was provided, and coordinate transformations were not performed. As such, both implementations were E($3$)-invariant.

\subsection*{Architecture of the multiscale Schake model}    
    Within the $l$th Schake message-passing layer (Figure \ref{Schake_diagram}A), the previous embedding from the $\left(l-1 \right)$th message-passing layer for the $i$th atom, $\bm{h}_i^{\left( l-1 \right)}$, and $\bm{h}_j^{\left( l-1 \right)}$ for all $j$ neighboring atoms within distance $r_{ji} \in \left[0, r^\mathrm{SAKE}_\mathrm{cut} \right)$ nm from atom $i$, are fed into a SAKE layer to compute a message $\bm{m}_i^{\left( l \right)}$. 
    The intermediate updated embedding for atom $i$, $\bm{\mathfrak{H}}_i^{\left( l \right)} = \bm{h}_i^{\left( l-1 \right)} + \bm{m}_i^{\left( l \right)}$, is then computed. 
    Next, $\bm{\mathfrak{H}}_i^{\left( l \right)}$ and $\bm{\mathfrak{H}}_k^{\left( l \right)}$ 
    for all $k$ neighboring alpha carbon atoms within distance $r_{ki} \in \left[r_\mathrm{cut}^\mathrm{SAKE}, r_\mathrm{cut}^\mathrm{SchNet} \right]$ nm from atom $i$ are fed into a SchNet layer to compute a message $\bm{v}_i^{\left( l \right)}$. The embedding is updated again to yield the embedding $\bm{h}_i^{\left( l \right)} = \bm{\mathfrak{H}}_i^{\left( l \right)} + \bm{v}_i^{\left( l \right)}$ for atom $i$. This embedding $\bm{h}_i^{\left( l \right)}$ is then either passed into another Schake layer, or used to compute the energy contribution of the $i$th atom. A detailed description of the Schake architecture is provided in the Supporting Information.

\subsection*{Training on the DISPEF subsets}
    PyTorch Distributed Data Parallel (DDP) was used to train all models across 8 Nvidia V100 GPUs. 
    All GPUs were provided and managed by the Massachusetts Institute of Technology (MIT) Supercloud and Lincoln Laboratory Supercomputing Center.\cite{reuther2018interactive}
    With the exception of ANI, all models were first trained for 140 epochs on DISPEF-S with an initial learning rate of $1 \times 10^{-3}$ that decayed every 3 epochs by a factor of 0.875. Next, the parameters obtained from training on DISPEF-S were used to initialize training on DISPEF-M. All models (except for ANI) were trained for 140 epochs on DISPEF-M with an initial learning rate of $1 \times 10^{-3}$ that decayed every 3 epochs by a factor of 0.875. The Adam optimizer and single-precision floating point format (FP32) were used for all training. The MAE per atom loss function, $\ell_\mathrm{MAE}^{E\mathrm{/a}} = \frac{1}{N} \sum_i^N \frac{1}{n_{\mathrm{atoms},i}}|E_\mathrm{pred}^i - E_\mathrm{target}^i |$, was used to train all models. Additionally, L2 regularization, $\ell_\mathrm{reg} = \lambda_\mathrm{reg} \sum_i \theta_i^2$, acted on the model parameters $\bm{\theta}$, with $\lambda_\mathrm{reg} = 1 \times 10^{-6}$. Regularization was implemented through the weight decay option when initializing the Adam optimizer. In total, the loss function was as follows: $\ell_\mathrm{tot} = \ell_\mathrm{MAE}^{E\mathrm{/a}} + \ell_\mathrm{reg}$. A custom, variable-size batch sampler was used when training all models except ANI, with the atom limit and maximum batch size set to maximize GPU memory utilization.

    For the ANI model, training for 860 epochs was first conducted on DISPEF-S with an initial learning rate of of $1 \times 10^{-3}$ that decayed every 3 epochs by a factor of 0.875. After epochs 210, 420, and 640, the learning rate and decay factor were set to $1 \times 10^{-3}$ and 0.875, $1 \times 10^{-4}$ and 0.9065, and $5 \times 10^{-4}$ and 0.886, respectively. Next, the parameters obtained from training on DISPEF-S were then used to initialize training on DISPEF-M. Training on DISPEF-M was conducted for 630 epochs. After epochs 210 and 420, the learning rate and decay rates were set to $1 \times 10^{-4}$ and 0.9065, and $5 \times 10^{-5}$ and 0.915, respectively.
    Following \citet{smith_less_2018}, both the AdamW\cite{loshchilov_decoupled_2019} and stochastic gradient descent\cite{robbins_stochastic_1951} (SGD) optimizers were used to train ANI. The AdamW optimizer was applied to all of the weights of ANI, while the SGD optimizer was applied to all of the bias parameters.
    Regularization was only applied to the weights of the 2$^\mathrm{nd}$ and 3$^\mathrm{rd}$ layers of each feed-forward NN, where $\lambda_\mathrm{reg} = 1 \times 10^{-5}$ for the 2$^\mathrm{nd}$ layer weights and $\lambda_\mathrm{reg} = 1 \times 10^{-6}$ for the 3$^\mathrm{rd}$ layer weights.

    More computationally-demanding force and energy training on DISPEF-M was also conducted for the SchNet, GN, SAKE 8S, SAKE 4M, Schake S, and Schake M architectures. Training was initialized using the optimal model parameters obtained from training these architectures to fit the energies of DISPEF-M. Training was conducted for 90 epochs with an initial learning rate of $1 \times 10^{-3}$ that decayed every 3 epochs by a factor of 0.875.
    Optimizers were applied following the same schemes used when training to fit only energies. The MAE per atom loss function was applied to the energies, while the MAE loss function, $\ell_\mathrm{MAE}^f = \frac{1}{N} \frac{1}{3n_\mathrm{atoms}} \sum_i^N \sum_j^{3n_\mathrm{atoms}} |f^{i,j}_\mathrm{pred} - f^{i,j}_\mathrm{target} |$, was applied to the forces. In total, the loss function was as follows: $\ell_\mathrm{tot} = \lambda_E \ell_\mathrm{MAE}^{E\mathrm{/a}} + \lambda_f \ell_\mathrm{MAE}^f$, where $\lambda_E = \lambda_f = 0.5$. 
    Three separate trainings with different random seeds were conducted, and the results for the best performing model are shown in Figure \ref{force_errors}.
    
    For ANI and for all of the GNNs explored in this study, the optimal hyperparameters that we selected are displayed in Tables S2 through S9. Additionally, a thorough analysis of computational cost (i.e., inference times and memory allocation) is presented in the Supporting Information, along with Figures S3 through S6 (smoothed using the LOWESS\cite{cleveland_robust_1979} algorithm for ease of interpretation) that show both inference times and memory allocations as a function of protein size.

\subsection*{Implementation of a variable-size batch sampler}
    
    While we initially trained our models on the DISPEF subsets using a standard, fixed-size PyTorch batch sampler, we found this to be computationally inefficient. Larger proteins contain more atoms / graph nodes, and thus more VRAM is required for training. While small molecules in datasets like MD17\cite{chmiela_machine_2017} (as one example) do not differ in size that greatly from one another, DISPEF-M contains proteins ranging from 16 to 390 amino acids. This vast difference in protein size renders a fixed-size batch sampler highly computationally inefficient. Even after shuffling the data, it is possible that any given batch will contain only smaller or larger proteins. The memory requirements for a batch containing only larger proteins will be significantly greater than that for a batch containing only smaller proteins. 
    As such, when using a fixed-size batch sampler, the batch size is limited by the size of the largest structures in the dataset. This leaves GPU memory left unused when processing batches containing smaller structures.
    
    To eliminate this inefficiency, we created a variable-size batch sampler. When constructing a batch, our variable-size batch sampler will compute a running summation of the total number of atoms in all proteins added to that batch. If adding the next protein to the current batch would result in the total atom number count being greater than a set limit to the number of atoms per batch, the next protein would be added to a new batch. 
    We found that this variable-size batch sampler results in a speed-up in training time of up to 25\%. 
    It is worth noting that this batch sampler introduces some complications when training a model across multiple GPUs using DDP. 
    We discuss these complications in-depth in the Supporting Information and show in Figure S7 that they don't adversely impact model accuracy.

\section{Supporting Information}
Description of the MAAPE relative error metric, our PyTorch implementation of SAKE, the Schake architecture, initial hyperparameter optimization, discussion on computational cost and our variable-sized batch sampler, along with additional supporting figures and tables.

\section{Data Avaiability}
Code for our PyTorch implementation of SAKE, our Schake architecture, and our variable-sized batch sampler will be made available on GitHub at the following link: \url{https://github.com/ZhangGroup-MITChemistry/Schake_GNN/}. We also include examples of how to use our implementation of SAKE, Schake, and our variable-sized batch sampler. Additionally, DISPEF-c and an example of how to load it will be made available on the GitHub repository linked above.
DISPEF-S, DISPEF-M, and DISPEF-L will be deposited to a suitable data hosting website.

\begin{acknowledgement}

    This work was supported by the National Institutes of Health (Grant R35GM133580). 
    J.A. was the recipient of the MIT Dean of Science Fellowship and the National Science Foundation Graduate Research Fellowship. Additionally, we thank undergraduate students Jack H. Liu from the MIT Department of Electrical Engineering and Computer Science and Chenxi Ye from the Peking University College of Chemistry and Molecular Engineering for their assistance implementing the ANI architecture using torchANI.
    We also thank graduate student Ivan Riveros from the MIT Department of Chemistry for his insight on GPU memory bandwidth.

\noindent\textbf{Competing interests:} 
Authors declare that they have no competing interests.

\end{acknowledgement}

\section{TOC Graphic}

\begin{figure}[H]
    \centering
    \includegraphics[width=1\textwidth]{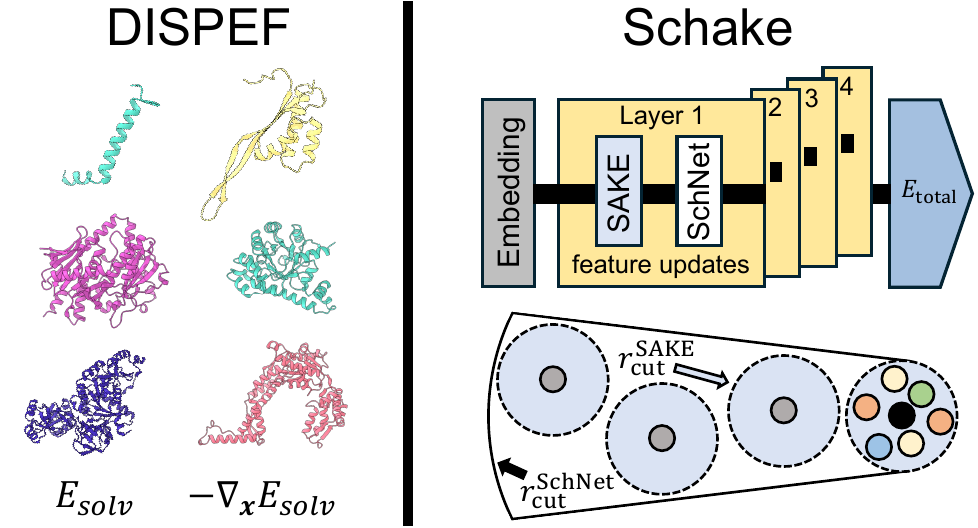}
    \label{TOC}
\end{figure}





\clearpage
\newpage
\bibliography{main}

\providecommand{\latin}[1]{#1}
\makeatletter
\providecommand{\doi}
  {\begingroup\let\do\@makeother\dospecials
  \catcode`\{=1 \catcode`\}=2 \doi@aux}
\providecommand{\doi@aux}[1]{\endgroup\texttt{#1}}
\makeatother
\providecommand*\mcitethebibliography{\thebibliography}
\csname @ifundefined\endcsname{endmcitethebibliography}  {\let\endmcitethebibliography\endthebibliography}{}
\begin{mcitethebibliography}{96}
\providecommand*\natexlab[1]{#1}
\providecommand*\mciteSetBstSublistMode[1]{}
\providecommand*\mciteSetBstMaxWidthForm[2]{}
\providecommand*\mciteBstWouldAddEndPuncttrue
  {\def\EndOfBibitem{\unskip.}}
\providecommand*\mciteBstWouldAddEndPunctfalse
  {\let\EndOfBibitem\relax}
\providecommand*\mciteSetBstMidEndSepPunct[3]{}
\providecommand*\mciteSetBstSublistLabelBeginEnd[3]{}
\providecommand*\EndOfBibitem{}
\mciteSetBstSublistMode{f}
\mciteSetBstMaxWidthForm{subitem}{(\alph{mcitesubitemcount})}
\mciteSetBstSublistLabelBeginEnd
  {\mcitemaxwidthsubitemform\space}
  {\relax}
  {\relax}

\bibitem[Airas \latin{et~al.}(2023)Airas, Ding, and Zhang]{airas_transferable_2023}
Airas,~J.; Ding,~X.; Zhang,~B. Transferable Implicit Solvation via Contrastive Learning of Graph Neural Networks. \emph{{ACS} Cent. Sci.} \textbf{2023}, \emph{9}, 2286--2297\relax
\mciteBstWouldAddEndPuncttrue
\mciteSetBstMidEndSepPunct{\mcitedefaultmidpunct}
{\mcitedefaultendpunct}{\mcitedefaultseppunct}\relax
\EndOfBibitem
\bibitem[Majewski \latin{et~al.}(2023)Majewski, Pérez, Thölke, Doerr, Charron, Giorgino, Husic, Clementi, Noé, and De~Fabritiis]{majewski_machine_2023}
Majewski,~M.; Pérez,~A.; Thölke,~P.; Doerr,~S.; Charron,~N.~E.; Giorgino,~T.; Husic,~B.~E.; Clementi,~C.; Noé,~F.; De~Fabritiis,~G. Machine learning coarse-grained potentials of protein thermodynamics. \emph{Nat. Commun.} \textbf{2023}, \emph{14}, 5739\relax
\mciteBstWouldAddEndPuncttrue
\mciteSetBstMidEndSepPunct{\mcitedefaultmidpunct}
{\mcitedefaultendpunct}{\mcitedefaultseppunct}\relax
\EndOfBibitem
\bibitem[Katzberger and Riniker(2023)Katzberger, and Riniker]{katzberger_implicit_2023}
Katzberger,~P.; Riniker,~S. Implicit solvent approach based on generalized Born and transferable graph neural networks for molecular dynamics simulations. \emph{J. Chem. Phys.} \textbf{2023}, \emph{158}, 204101\relax
\mciteBstWouldAddEndPuncttrue
\mciteSetBstMidEndSepPunct{\mcitedefaultmidpunct}
{\mcitedefaultendpunct}{\mcitedefaultseppunct}\relax
\EndOfBibitem
\bibitem[Katzberger and Riniker(2024)Katzberger, and Riniker]{katzberger_general_2024}
Katzberger,~P.; Riniker,~S. A general graph neural network based implicit solvation model for organic molecules in water. \emph{Chem. Sci.} \textbf{2024}, \relax
\mciteBstWouldAddEndPunctfalse
\mciteSetBstMidEndSepPunct{\mcitedefaultmidpunct}
{}{\mcitedefaultseppunct}\relax
\EndOfBibitem
\bibitem[Chen \latin{et~al.}(2021)Chen, Krämer, Charron, Husic, Clementi, and Noé]{chen_machine_2021}
Chen,~Y.; Krämer,~A.; Charron,~N.~E.; Husic,~B.~E.; Clementi,~C.; Noé,~F. Machine learning implicit solvation for molecular dynamics. \emph{J. Chem. Phys.} \textbf{2021}, \emph{155}, 084101\relax
\mciteBstWouldAddEndPuncttrue
\mciteSetBstMidEndSepPunct{\mcitedefaultmidpunct}
{\mcitedefaultendpunct}{\mcitedefaultseppunct}\relax
\EndOfBibitem
\bibitem[Husic \latin{et~al.}(2020)Husic, Charron, Lemm, Wang, Pérez, Majewski, Krämer, Chen, Olsson, de~Fabritiis, Noé, and Clementi]{husic_coarse_2020}
Husic,~B.~E.; Charron,~N.~E.; Lemm,~D.; Wang,~J.; Pérez,~A.; Majewski,~M.; Krämer,~A.; Chen,~Y.; Olsson,~S.; de~Fabritiis,~G. \latin{et~al.}  Coarse graining molecular dynamics with graph neural networks. \emph{J. Chem. Phys.} \textbf{2020}, \emph{153}, 194101\relax
\mciteBstWouldAddEndPuncttrue
\mciteSetBstMidEndSepPunct{\mcitedefaultmidpunct}
{\mcitedefaultendpunct}{\mcitedefaultseppunct}\relax
\EndOfBibitem
\bibitem[Kozinsky \latin{et~al.}(2023)Kozinsky, Musaelian, Johansson, and Batzner]{kozinsky_scaling_2023}
Kozinsky,~B.; Musaelian,~A.; Johansson,~A.; Batzner,~S. Scaling the Leading Accuracy of Deep Equivariant Models to Biomolecular Simulations of Realistic Size. Proceedings of the International Conference for High Performance Computing, Networking, Storage and Analysis. 2023; pp 1--12\relax
\mciteBstWouldAddEndPuncttrue
\mciteSetBstMidEndSepPunct{\mcitedefaultmidpunct}
{\mcitedefaultendpunct}{\mcitedefaultseppunct}\relax
\EndOfBibitem
\bibitem[Wang \latin{et~al.}(2023)Wang, Wu, Sun, He, Liu, Shao, Wang, and Liu]{wang_improving_2023}
Wang,~Z.; Wu,~H.; Sun,~L.; He,~X.; Liu,~Z.; Shao,~B.; Wang,~T.; Liu,~T.-Y. Improving machine learning force fields for molecular dynamics simulations with fine-grained force metrics. \emph{J. Chem. Phys.} \textbf{2023}, \emph{159}, 035101\relax
\mciteBstWouldAddEndPuncttrue
\mciteSetBstMidEndSepPunct{\mcitedefaultmidpunct}
{\mcitedefaultendpunct}{\mcitedefaultseppunct}\relax
\EndOfBibitem
\bibitem[Brunken \latin{et~al.}(2024)Brunken, Boyer, Omar, Diallo, Beguir, Carranza, and Bent]{brunken_machine_2024}
Brunken,~C.; Boyer,~S.; Omar,~M.; Diallo,~B.~N.; Beguir,~K.; Carranza,~N.~L.; Bent,~O. Machine learning of force fields towards molecular dynamics simulations of proteins at {DFT} accuracy. \emph{ICLR 2024 Workshop on Generative and Experimental Perspectives for Biomolecular Design}. Vienna, Austria. 2024\relax
\mciteBstWouldAddEndPuncttrue
\mciteSetBstMidEndSepPunct{\mcitedefaultmidpunct}
{\mcitedefaultendpunct}{\mcitedefaultseppunct}\relax
\EndOfBibitem
\bibitem[Yao \latin{et~al.}(2023)Yao, Van, Pan, Park, Mao, Pu, Mei, and Shao]{yao_machine_2023}
Yao,~S.; Van,~R.; Pan,~X.; Park,~J.~H.; Mao,~Y.; Pu,~J.; Mei,~Y.; Shao,~Y. Machine learning based implicit solvent model for aqueous-solution alanine dipeptide molecular dynamics simulations. \emph{{RSC} Adv.} \textbf{2023}, \emph{13}, 4565--4577\relax
\mciteBstWouldAddEndPuncttrue
\mciteSetBstMidEndSepPunct{\mcitedefaultmidpunct}
{\mcitedefaultendpunct}{\mcitedefaultseppunct}\relax
\EndOfBibitem
\bibitem[Wang \latin{et~al.}(2019)Wang, Olsson, Wehmeyer, Pérez, Charron, de~Fabritiis, Noé, and Clementi]{wang_machine_2019}
Wang,~J.; Olsson,~S.; Wehmeyer,~C.; Pérez,~A.; Charron,~N.~E.; de~Fabritiis,~G.; Noé,~F.; Clementi,~C. Machine Learning of Coarse-Grained Molecular Dynamics Force Fields. \emph{{ACS} Cent. Sci.} \textbf{2019}, \emph{5}, 755--767\relax
\mciteBstWouldAddEndPuncttrue
\mciteSetBstMidEndSepPunct{\mcitedefaultmidpunct}
{\mcitedefaultendpunct}{\mcitedefaultseppunct}\relax
\EndOfBibitem
\bibitem[Wang \latin{et~al.}(2021)Wang, Charron, Husic, Olsson, Noé, and Clementi]{wang_multi-body_2021}
Wang,~J.; Charron,~N.; Husic,~B.; Olsson,~S.; Noé,~F.; Clementi,~C. Multi-body effects in a coarse-grained protein force field. \emph{J. Chem. Phys.} \textbf{2021}, \emph{154}, 164113\relax
\mciteBstWouldAddEndPuncttrue
\mciteSetBstMidEndSepPunct{\mcitedefaultmidpunct}
{\mcitedefaultendpunct}{\mcitedefaultseppunct}\relax
\EndOfBibitem
\bibitem[Fu \latin{et~al.}(2023)Fu, Wu, Wang, Xie, Keten, Gomez-Bombarelli, and Jaakkola]{fu_forces_2023}
Fu,~X.; Wu,~Z.; Wang,~W.; Xie,~T.; Keten,~S.; Gomez-Bombarelli,~R.; Jaakkola,~T.~S. Forces are not Enough: Benchmark and Critical Evaluation for Machine Learning Force Fields with Molecular Simulations. \emph{{arXiv}} \textbf{2023}, {arXiv}:2210.07237, accessed on 2024-04-09\relax
\mciteBstWouldAddEndPuncttrue
\mciteSetBstMidEndSepPunct{\mcitedefaultmidpunct}
{\mcitedefaultendpunct}{\mcitedefaultseppunct}\relax
\EndOfBibitem
\bibitem[Durumeric \latin{et~al.}(2024)Durumeric, Chen, Noé, and Clementi]{durumeric_learning_2024}
Durumeric,~A. E.~P.; Chen,~Y.; Noé,~F.; Clementi,~C. Learning data efficient coarse-grained molecular dynamics from forces and noise. \emph{{arXiv}} \textbf{2024}, {arXiv}:2407.01286, accessed on 2024-09-06\relax
\mciteBstWouldAddEndPuncttrue
\mciteSetBstMidEndSepPunct{\mcitedefaultmidpunct}
{\mcitedefaultendpunct}{\mcitedefaultseppunct}\relax
\EndOfBibitem
\bibitem[Arts \latin{et~al.}(2023)Arts, Garcia~Satorras, Huang, Zügner, Federici, Clementi, Noé, Pinsler, and van~den Berg]{arts_two_2023}
Arts,~M.; Garcia~Satorras,~V.; Huang,~C.-W.; Zügner,~D.; Federici,~M.; Clementi,~C.; Noé,~F.; Pinsler,~R.; van~den Berg,~R. Two for One: Diffusion Models and Force Fields for Coarse-Grained Molecular Dynamics. \emph{J. Chem. Theory Comput.} \textbf{2023}, \emph{19}, 6151--6159\relax
\mciteBstWouldAddEndPuncttrue
\mciteSetBstMidEndSepPunct{\mcitedefaultmidpunct}
{\mcitedefaultendpunct}{\mcitedefaultseppunct}\relax
\EndOfBibitem
\bibitem[Köhler \latin{et~al.}(2023)Köhler, Chen, Krämer, Clementi, and Noé]{kohler_flow-matching_2023}
Köhler,~J.; Chen,~Y.; Krämer,~A.; Clementi,~C.; Noé,~F. Flow-Matching: Efficient Coarse-Graining of Molecular Dynamics without Forces. \emph{J. Chem. Theory Comput.} \textbf{2023}, \emph{19}, 942--952\relax
\mciteBstWouldAddEndPuncttrue
\mciteSetBstMidEndSepPunct{\mcitedefaultmidpunct}
{\mcitedefaultendpunct}{\mcitedefaultseppunct}\relax
\EndOfBibitem
\bibitem[Loose \latin{et~al.}(2023)Loose, Sahrmann, Qu, and Voth]{loose_coarse-graining_2023}
Loose,~T.~D.; Sahrmann,~P.~G.; Qu,~T.~S.; Voth,~G.~A. Coarse-Graining with Equivariant Neural Networks: A Path Toward Accurate and Data-Efficient Models. \emph{J. Phys. Chem. B} \textbf{2023}, \emph{127}, 10564--10572\relax
\mciteBstWouldAddEndPuncttrue
\mciteSetBstMidEndSepPunct{\mcitedefaultmidpunct}
{\mcitedefaultendpunct}{\mcitedefaultseppunct}\relax
\EndOfBibitem
\bibitem[Durumeric and Voth(2019)Durumeric, and Voth]{durumeric_adversarial-residual-coarse-graining_2019}
Durumeric,~A. E.~P.; Voth,~G.~A. Adversarial-residual-coarse-graining: Applying machine learning theory to systematic molecular coarse-graining. \emph{J. Chem. Phys.} \textbf{2019}, \emph{151}, 124110\relax
\mciteBstWouldAddEndPuncttrue
\mciteSetBstMidEndSepPunct{\mcitedefaultmidpunct}
{\mcitedefaultendpunct}{\mcitedefaultseppunct}\relax
\EndOfBibitem
\bibitem[Durumeric \latin{et~al.}(2023)Durumeric, Charron, Templeton, Musil, Bonneau, Pasos-Trejo, Chen, Kelkar, Noé, and Clementi]{durumeric_machine_2023}
Durumeric,~A. E.~P.; Charron,~N.~E.; Templeton,~C.; Musil,~F.; Bonneau,~K.; Pasos-Trejo,~A.~S.; Chen,~Y.; Kelkar,~A.; Noé,~F.; Clementi,~C. Machine learned coarse-grained protein force-fields: Are we there yet? \emph{Curr. Opin. Struct. Biol.} \textbf{2023}, \emph{79}, 102533\relax
\mciteBstWouldAddEndPuncttrue
\mciteSetBstMidEndSepPunct{\mcitedefaultmidpunct}
{\mcitedefaultendpunct}{\mcitedefaultseppunct}\relax
\EndOfBibitem
\bibitem[Ding and Zhang(2022)Ding, and Zhang]{ding_contrastive_2022}
Ding,~X.; Zhang,~B. Contrastive Learning of Coarse-Grained Force Fields. \emph{J. Chem. Theory Comput.} \textbf{2022}, \emph{18}, 6334--6344\relax
\mciteBstWouldAddEndPuncttrue
\mciteSetBstMidEndSepPunct{\mcitedefaultmidpunct}
{\mcitedefaultendpunct}{\mcitedefaultseppunct}\relax
\EndOfBibitem
\bibitem[Ding(2024)]{ding_optimizing_2024}
Ding,~X. Optimizing Force Fields with Experimental Data Using Ensemble Reweighting and Potential Contrasting. \emph{{ChemRxiv}} \textbf{2024}, chemrxiv--2024--lvlwb, accessed on 2024-04-09\relax
\mciteBstWouldAddEndPuncttrue
\mciteSetBstMidEndSepPunct{\mcitedefaultmidpunct}
{\mcitedefaultendpunct}{\mcitedefaultseppunct}\relax
\EndOfBibitem
\bibitem[Chmiela \latin{et~al.}(2018)Chmiela, Sauceda, Müller, and Tkatchenko]{chmiela_towards_2018}
Chmiela,~S.; Sauceda,~H.~E.; Müller,~K.-R.; Tkatchenko,~A. Towards exact molecular dynamics simulations with machine-learned force fields. \emph{Nat. Commun.} \textbf{2018}, \emph{9}, 3887\relax
\mciteBstWouldAddEndPuncttrue
\mciteSetBstMidEndSepPunct{\mcitedefaultmidpunct}
{\mcitedefaultendpunct}{\mcitedefaultseppunct}\relax
\EndOfBibitem
\bibitem[Chmiela \latin{et~al.}(2020)Chmiela, Sauceda, Tkatchenko, and Müller]{chmiela_accurate_2020}
Chmiela,~S.; Sauceda,~H.~E.; Tkatchenko,~A.; Müller,~K.-R. In \emph{Machine Learning Meets Quantum Physics}; Schütt,~K.~T., Chmiela,~S., von Lilienfeld,~O.~A., Tkatchenko,~A., Tsuda,~K., Müller,~K.-R., Eds.; Springer International Publishing, 2020; pp 129--154\relax
\mciteBstWouldAddEndPuncttrue
\mciteSetBstMidEndSepPunct{\mcitedefaultmidpunct}
{\mcitedefaultendpunct}{\mcitedefaultseppunct}\relax
\EndOfBibitem
\bibitem[Duschatko \latin{et~al.}(2024)Duschatko, Fu, Owen, Xie, Musaelian, Jaakkola, and Kozinsky]{duschatko_thermodynamically_2024}
Duschatko,~B.~R.; Fu,~X.; Owen,~C.; Xie,~Y.; Musaelian,~A.; Jaakkola,~T.; Kozinsky,~B. Thermodynamically Informed Multimodal Learning of High-Dimensional Free Energy Models in Molecular Coarse Graining. \emph{{arXiv}} \textbf{2024}, {arXiv}:2405.19386, accessed on 2024-06-24\relax
\mciteBstWouldAddEndPuncttrue
\mciteSetBstMidEndSepPunct{\mcitedefaultmidpunct}
{\mcitedefaultendpunct}{\mcitedefaultseppunct}\relax
\EndOfBibitem
\bibitem[G. Greener(2024)]{ggreener_differentiable_2024}
G. Greener,~J. Differentiable simulation to develop molecular dynamics force fields for disordered proteins. \emph{Chem. Sci.} \textbf{2024}, \emph{15}, 4897--4909\relax
\mciteBstWouldAddEndPuncttrue
\mciteSetBstMidEndSepPunct{\mcitedefaultmidpunct}
{\mcitedefaultendpunct}{\mcitedefaultseppunct}\relax
\EndOfBibitem
\bibitem[Yang \latin{et~al.}(2023)Yang, Templeton, Rosenberger, Bittracher, Nüske, Noé, and Clementi]{yang_slicing_2023}
Yang,~W.; Templeton,~C.; Rosenberger,~D.; Bittracher,~A.; Nüske,~F.; Noé,~F.; Clementi,~C. Slicing and Dicing: Optimal Coarse-Grained Representation to Preserve Molecular Kinetics. \emph{{ACS} Cent. Sci.} \textbf{2023}, \emph{9}, 186--196\relax
\mciteBstWouldAddEndPuncttrue
\mciteSetBstMidEndSepPunct{\mcitedefaultmidpunct}
{\mcitedefaultendpunct}{\mcitedefaultseppunct}\relax
\EndOfBibitem
\bibitem[Corso \latin{et~al.}(2024)Corso, Stark, Jegelka, Jaakkola, and Barzilay]{corso_graph_2024}
Corso,~G.; Stark,~H.; Jegelka,~S.; Jaakkola,~T.; Barzilay,~R. Graph neural networks. \emph{Nat. Rev. Methods Primers} \textbf{2024}, \emph{4}, 1--13\relax
\mciteBstWouldAddEndPuncttrue
\mciteSetBstMidEndSepPunct{\mcitedefaultmidpunct}
{\mcitedefaultendpunct}{\mcitedefaultseppunct}\relax
\EndOfBibitem
\bibitem[Batzner \latin{et~al.}(2023)Batzner, Musaelian, and Kozinsky]{batzner_advancing_2023}
Batzner,~S.; Musaelian,~A.; Kozinsky,~B. Advancing molecular simulation with equivariant interatomic potentials. \emph{Nat. Rev. Phys.} \textbf{2023}, \emph{5}, 437--438\relax
\mciteBstWouldAddEndPuncttrue
\mciteSetBstMidEndSepPunct{\mcitedefaultmidpunct}
{\mcitedefaultendpunct}{\mcitedefaultseppunct}\relax
\EndOfBibitem
\bibitem[Eastman \latin{et~al.}(2024)Eastman, Galvelis, Peláez, Abreu, Farr, Gallicchio, Gorenko, Henry, Hu, Huang, Krämer, Michel, Mitchell, Pande, Rodrigues, Rodriguez-Guerra, Simmonett, Singh, Swails, Turner, Wang, Zhang, Chodera, De~Fabritiis, and Markland]{eastman_openmm_2024}
Eastman,~P.; Galvelis,~R.; Peláez,~R.~P.; Abreu,~C. R.~A.; Farr,~S.~E.; Gallicchio,~E.; Gorenko,~A.; Henry,~M.~M.; Hu,~F.; Huang,~J. \latin{et~al.}  {OpenMM} 8: Molecular Dynamics Simulation with Machine Learning Potentials. \emph{J. Phys. Chem. B} \textbf{2024}, \emph{128}, 109--116\relax
\mciteBstWouldAddEndPuncttrue
\mciteSetBstMidEndSepPunct{\mcitedefaultmidpunct}
{\mcitedefaultendpunct}{\mcitedefaultseppunct}\relax
\EndOfBibitem
\bibitem[Zheng \latin{et~al.}(2024)Zheng, He, Liu, Shi, Lu, Feng, Ju, Wang, Zhu, Min, Zhang, Tang, Hao, Jin, Chen, Noé, Liu, and Liu]{zheng_predicting_2024}
Zheng,~S.; He,~J.; Liu,~C.; Shi,~Y.; Lu,~Z.; Feng,~W.; Ju,~F.; Wang,~J.; Zhu,~J.; Min,~Y. \latin{et~al.}  Predicting equilibrium distributions for molecular systems with deep learning. \emph{Nat. Mach. Intell.} \textbf{2024}, \emph{6}, 558--567\relax
\mciteBstWouldAddEndPuncttrue
\mciteSetBstMidEndSepPunct{\mcitedefaultmidpunct}
{\mcitedefaultendpunct}{\mcitedefaultseppunct}\relax
\EndOfBibitem
\bibitem[Sahrmann \latin{et~al.}(2023)Sahrmann, Loose, Durumeric, and Voth]{sahrmann_utilizing_2023}
Sahrmann,~P.~G.; Loose,~T.~D.; Durumeric,~A. E.~P.; Voth,~G.~A. Utilizing Machine Learning to Greatly Expand the Range and Accuracy of Bottom-Up Coarse-Grained Models through Virtual Particles. \emph{J. Chem. Theory Comput.} \textbf{2023}, \emph{19}, 4402--4413\relax
\mciteBstWouldAddEndPuncttrue
\mciteSetBstMidEndSepPunct{\mcitedefaultmidpunct}
{\mcitedefaultendpunct}{\mcitedefaultseppunct}\relax
\EndOfBibitem
\bibitem[Bonneau \latin{et~al.}(2024)Bonneau, Lederer, Templeton, Rosenberger, Müller, and Clementi]{bonneau_peering_2024}
Bonneau,~K.; Lederer,~J.; Templeton,~C.; Rosenberger,~D.; Müller,~K.-R.; Clementi,~C. Peering inside the black box: Learning the relevance of many-body functions in Neural Network potentials. \emph{{arXiv}} \textbf{2024}, {arXiv}:2407.04526, accessed on 2024-09-06\relax
\mciteBstWouldAddEndPuncttrue
\mciteSetBstMidEndSepPunct{\mcitedefaultmidpunct}
{\mcitedefaultendpunct}{\mcitedefaultseppunct}\relax
\EndOfBibitem
\bibitem[Anstine and Isayev(2023)Anstine, and Isayev]{anstine_machine_2023}
Anstine,~D.~M.; Isayev,~O. Machine Learning Interatomic Potentials and Long-Range Physics. \emph{J. Phys. Chem. A} \textbf{2023}, \emph{127}, 2417--2431\relax
\mciteBstWouldAddEndPuncttrue
\mciteSetBstMidEndSepPunct{\mcitedefaultmidpunct}
{\mcitedefaultendpunct}{\mcitedefaultseppunct}\relax
\EndOfBibitem
\bibitem[Cheng \latin{et~al.}(2024)Cheng, Bi, Liu, Chen, and Yu]{cheng_developing_2024}
Cheng,~Z.; Bi,~H.; Liu,~S.; Chen,~J.; Yu,~K. Developing Differentiable Long-Range Force Field for Proteins with E(3) Neural Network Predicted Asymptotic Parameters. \emph{{ChemRxiv}} \textbf{2024}, chemrxiv--2024--kkb5h, accessed on 2024-04-09\relax
\mciteBstWouldAddEndPuncttrue
\mciteSetBstMidEndSepPunct{\mcitedefaultmidpunct}
{\mcitedefaultendpunct}{\mcitedefaultseppunct}\relax
\EndOfBibitem
\bibitem[Faller \latin{et~al.}(2024)Faller, Kaltak, and Kresse]{faller_density-based_2024}
Faller,~C.; Kaltak,~M.; Kresse,~G. Density-Based Long-Range Electrostatic Descriptors for Machine Learning Force Fields. \emph{{arXiv}} \textbf{2024}, {arXiv}:2406.17595, accessed on 2024-08-30\relax
\mciteBstWouldAddEndPuncttrue
\mciteSetBstMidEndSepPunct{\mcitedefaultmidpunct}
{\mcitedefaultendpunct}{\mcitedefaultseppunct}\relax
\EndOfBibitem
\bibitem[Takaba \latin{et~al.}(2024)Takaba, J. Friedman, E. Cavender, Kumar Behara, Pulido, M. Henry, {MacDermott}-Opeskin, R. Iacovella, M. Nagle, Matthew Payne, R. Shirts, L. Mobley, D. Chodera, and Wang]{takaba_machine-learned_2024}
Takaba,~K.; J. Friedman,~A.; E. Cavender,~C.; Kumar Behara,~P.; Pulido,~I.; M. Henry,~M.; {MacDermott}-Opeskin,~H.; R. Iacovella,~C.; M. Nagle,~A.; Matthew Payne,~A. \latin{et~al.}  Machine-learned molecular mechanics force fields from large-scale quantum chemical data. \emph{Chem. Sci.} \textbf{2024}, \emph{15}, 12861--12878\relax
\mciteBstWouldAddEndPuncttrue
\mciteSetBstMidEndSepPunct{\mcitedefaultmidpunct}
{\mcitedefaultendpunct}{\mcitedefaultseppunct}\relax
\EndOfBibitem
\bibitem[Wang \latin{et~al.}(2024)Wang, Pulido, Takaba, Kaminow, Scheen, Wang, and Chodera]{wang_espalomacharge_2024}
Wang,~Y.; Pulido,~I.; Takaba,~K.; Kaminow,~B.; Scheen,~J.; Wang,~L.; Chodera,~J.~D. {EspalomaCharge}: Machine Learning-Enabled Ultrafast Partial Charge Assignment. \emph{J. Phys. Chem. A} \textbf{2024}, \emph{128}, 4160--4167\relax
\mciteBstWouldAddEndPuncttrue
\mciteSetBstMidEndSepPunct{\mcitedefaultmidpunct}
{\mcitedefaultendpunct}{\mcitedefaultseppunct}\relax
\EndOfBibitem
\bibitem[Wang \latin{et~al.}(2022)Wang, Fass, Kaminow, E. Herr, Rufa, Zhang, Pulido, Henry, Macdonald, Takaba, and D. Chodera]{wang_end--end_2022}
Wang,~Y.; Fass,~J.; Kaminow,~B.; E. Herr,~J.; Rufa,~D.; Zhang,~I.; Pulido,~I.; Henry,~M.; Macdonald,~H. E.~B.; Takaba,~K. \latin{et~al.}  End-to-end differentiable construction of molecular mechanics force fields. \emph{Chem. Sci.} \textbf{2022}, \emph{13}, 12016--12033\relax
\mciteBstWouldAddEndPuncttrue
\mciteSetBstMidEndSepPunct{\mcitedefaultmidpunct}
{\mcitedefaultendpunct}{\mcitedefaultseppunct}\relax
\EndOfBibitem
\bibitem[Galvelis \latin{et~al.}(2023)Galvelis, Varela-Rial, Doerr, Fino, Eastman, Markland, Chodera, and De~Fabritiis]{galvelis_nnpmm_2023}
Galvelis,~R.; Varela-Rial,~A.; Doerr,~S.; Fino,~R.; Eastman,~P.; Markland,~T.~E.; Chodera,~J.~D.; De~Fabritiis,~G. {NNP}/{MM}: Accelerating Molecular Dynamics Simulations with Machine Learning Potentials and Molecular Mechanics. \emph{J. Chem. Inf. Model.} \textbf{2023}, \emph{63}, 5701--5708\relax
\mciteBstWouldAddEndPuncttrue
\mciteSetBstMidEndSepPunct{\mcitedefaultmidpunct}
{\mcitedefaultendpunct}{\mcitedefaultseppunct}\relax
\EndOfBibitem
\bibitem[Sabanés~Zariquiey \latin{et~al.}(2024)Sabanés~Zariquiey, Galvelis, Gallicchio, Chodera, Markland, and De~Fabritiis]{sabanes_zariquiey_enhancing_2024}
Sabanés~Zariquiey,~F.; Galvelis,~R.; Gallicchio,~E.; Chodera,~J.~D.; Markland,~T.~E.; De~Fabritiis,~G. Enhancing Protein–Ligand Binding Affinity Predictions Using Neural Network Potentials. \emph{J. Chem. Inf. Model.} \textbf{2024}, \emph{64}, 1481--1485\relax
\mciteBstWouldAddEndPuncttrue
\mciteSetBstMidEndSepPunct{\mcitedefaultmidpunct}
{\mcitedefaultendpunct}{\mcitedefaultseppunct}\relax
\EndOfBibitem
\bibitem[Barnett and Chodera(2024)Barnett, and Chodera]{barnett_neural_2024}
Barnett,~S.; Chodera,~J.~D. Neural Network Potentials for Enabling Advanced Small-Molecule Drug Discovery and Generative Design. \emph{{GEN} Biotechnology} \textbf{2024}, \emph{3}, 119--129\relax
\mciteBstWouldAddEndPuncttrue
\mciteSetBstMidEndSepPunct{\mcitedefaultmidpunct}
{\mcitedefaultendpunct}{\mcitedefaultseppunct}\relax
\EndOfBibitem
\bibitem[Unke and Meuwly(2019)Unke, and Meuwly]{unke_physnet_2019}
Unke,~O.~T.; Meuwly,~M. {PhysNet}: A Neural Network for Predicting Energies, Forces, Dipole Moments, and Partial Charges. \emph{J. Chem. Theory Comput.} \textbf{2019}, \emph{15}, 3678--3693\relax
\mciteBstWouldAddEndPuncttrue
\mciteSetBstMidEndSepPunct{\mcitedefaultmidpunct}
{\mcitedefaultendpunct}{\mcitedefaultseppunct}\relax
\EndOfBibitem
\bibitem[Gasteiger \latin{et~al.}(2022)Gasteiger, Groß, and Günnemann]{gasteiger_directional_2022}
Gasteiger,~J.; Groß,~J.; Günnemann,~S. Directional Message Passing for Molecular Graphs. \emph{{arXiv}} \textbf{2022}, {arXiv}:2003.03123, accessed on 2023-07-13\relax
\mciteBstWouldAddEndPuncttrue
\mciteSetBstMidEndSepPunct{\mcitedefaultmidpunct}
{\mcitedefaultendpunct}{\mcitedefaultseppunct}\relax
\EndOfBibitem
\bibitem[Gasteiger \latin{et~al.}(2022)Gasteiger, Becker, and Günnemann]{gasteiger_gemnet_2022}
Gasteiger,~J.; Becker,~F.; Günnemann,~S. {GemNet}: Universal Directional Graph Neural Networks for Molecules. \emph{{arXiv}} \textbf{2022}, {arXiv}:2106.08903, accessed on 2023-07-13\relax
\mciteBstWouldAddEndPuncttrue
\mciteSetBstMidEndSepPunct{\mcitedefaultmidpunct}
{\mcitedefaultendpunct}{\mcitedefaultseppunct}\relax
\EndOfBibitem
\bibitem[Anderson \latin{et~al.}(2019)Anderson, Hy, and Kondor]{anderson_cormorant_2019}
Anderson,~B.; Hy,~T.-S.; Kondor,~R. Cormorant: Covariant Molecular Neural Networks. \emph{{arXiv}} \textbf{2019}, {arXiv}:1906.04015, accessed on 2023-7-13\relax
\mciteBstWouldAddEndPuncttrue
\mciteSetBstMidEndSepPunct{\mcitedefaultmidpunct}
{\mcitedefaultendpunct}{\mcitedefaultseppunct}\relax
\EndOfBibitem
\bibitem[Brandstetter \latin{et~al.}(2022)Brandstetter, Hesselink, van~der Pol, Bekkers, and Welling]{brandstetter_geometric_2022}
Brandstetter,~J.; Hesselink,~R.; van~der Pol,~E.; Bekkers,~E.~J.; Welling,~M. Geometric and Physical Quantities Improve E(3) Equivariant Message Passing. \emph{{arXiv}} \textbf{2022}, {arXiv}:2110.02905, accessed on 2023-07-13\relax
\mciteBstWouldAddEndPuncttrue
\mciteSetBstMidEndSepPunct{\mcitedefaultmidpunct}
{\mcitedefaultendpunct}{\mcitedefaultseppunct}\relax
\EndOfBibitem
\bibitem[Li \latin{et~al.}(2022)Li, Liu, Wang, Du, and Chen]{li_egnn_2022}
Li,~Y.; Liu,~L.; Wang,~G.; Du,~Y.; Chen,~P. {EGNN}: Constructing explainable graph neural networks via knowledge distillation. \textbf{2022}, \emph{241}, 108345\relax
\mciteBstWouldAddEndPuncttrue
\mciteSetBstMidEndSepPunct{\mcitedefaultmidpunct}
{\mcitedefaultendpunct}{\mcitedefaultseppunct}\relax
\EndOfBibitem
\bibitem[Schütt \latin{et~al.}(2021)Schütt, Unke, and Gastegger]{schutt_equivariant_2021}
Schütt,~K.~T.; Unke,~O.~T.; Gastegger,~M. Equivariant message passing for the prediction of tensorial properties and molecular spectra. \emph{{arXiv}} \textbf{2021}, {arXiv}:2102.03150, accessed on 2023-07-13\relax
\mciteBstWouldAddEndPuncttrue
\mciteSetBstMidEndSepPunct{\mcitedefaultmidpunct}
{\mcitedefaultendpunct}{\mcitedefaultseppunct}\relax
\EndOfBibitem
\bibitem[Thomas \latin{et~al.}(2018)Thomas, Smidt, Kearnes, Yang, Li, Kohlhoff, and Riley]{thomas_tensor_2018}
Thomas,~N.; Smidt,~T.; Kearnes,~S.; Yang,~L.; Li,~L.; Kohlhoff,~K.; Riley,~P. Tensor field networks: Rotation- and translation-equivariant neural networks for 3D point clouds. \emph{{arXiv}} \textbf{2018}, {arXiv}:1802.08219, accessed on 2023-7-13\relax
\mciteBstWouldAddEndPuncttrue
\mciteSetBstMidEndSepPunct{\mcitedefaultmidpunct}
{\mcitedefaultendpunct}{\mcitedefaultseppunct}\relax
\EndOfBibitem
\bibitem[Fuchs \latin{et~al.}(2020)Fuchs, Worrall, Fischer, and Welling]{fuchs_se3-transformers_2020}
Fuchs,~F.~B.; Worrall,~D.~E.; Fischer,~V.; Welling,~M. {SE}(3)-Transformers: 3D Roto-Translation Equivariant Attention Networks. \emph{{arXiv}} \textbf{2020}, {arXiv}:2006.10503, accessed on 2023-07-13\relax
\mciteBstWouldAddEndPuncttrue
\mciteSetBstMidEndSepPunct{\mcitedefaultmidpunct}
{\mcitedefaultendpunct}{\mcitedefaultseppunct}\relax
\EndOfBibitem
\bibitem[Huang \latin{et~al.}(2022)Huang, Han, Rong, Xu, Sun, and Huang]{huang_equivariant_2022}
Huang,~W.; Han,~J.; Rong,~Y.; Xu,~T.; Sun,~F.; Huang,~J. Equivariant Graph Mechanics Networks with Constraints. \emph{{arXiv}} \textbf{2022}, {arXiv}:2203.06442, accessed on 2023-07-19\relax
\mciteBstWouldAddEndPuncttrue
\mciteSetBstMidEndSepPunct{\mcitedefaultmidpunct}
{\mcitedefaultendpunct}{\mcitedefaultseppunct}\relax
\EndOfBibitem
\bibitem[Schütt \latin{et~al.}(2018)Schütt, Sauceda, Kindermans, Tkatchenko, and Müller]{schutt_schnet_2018}
Schütt,~K.~T.; Sauceda,~H.~E.; Kindermans,~P.-J.; Tkatchenko,~A.; Müller,~K.-R. {SchNet} – A deep learning architecture for molecules and materials. \emph{J. Chem. Phys.} \textbf{2018}, \emph{148}, 241722\relax
\mciteBstWouldAddEndPuncttrue
\mciteSetBstMidEndSepPunct{\mcitedefaultmidpunct}
{\mcitedefaultendpunct}{\mcitedefaultseppunct}\relax
\EndOfBibitem
\bibitem[Satorras \latin{et~al.}(2022)Satorras, Hoogeboom, and Welling]{satorras_en_2022}
Satorras,~V.~G.; Hoogeboom,~E.; Welling,~M. E(n) Equivariant Graph Neural Networks. \emph{{arXiv}} \textbf{2022}, {arXiv}:2102.09844, accessed on 2023-07-19\relax
\mciteBstWouldAddEndPuncttrue
\mciteSetBstMidEndSepPunct{\mcitedefaultmidpunct}
{\mcitedefaultendpunct}{\mcitedefaultseppunct}\relax
\EndOfBibitem
\bibitem[Wang and Chodera(2023)Wang, and Chodera]{wang_spatial_2023}
Wang,~Y.; Chodera,~J.~D. Spatial Attention Kinetic Networks with E(n)-Equivariance. \emph{{arXiv}} \textbf{2023}, {arXiv}:2301.08893, accessed on 2023-07-13\relax
\mciteBstWouldAddEndPuncttrue
\mciteSetBstMidEndSepPunct{\mcitedefaultmidpunct}
{\mcitedefaultendpunct}{\mcitedefaultseppunct}\relax
\EndOfBibitem
\bibitem[Thölke and de~Fabritiis(2022)Thölke, and de~Fabritiis]{tholke_torchmd-net_2022}
Thölke,~P.; de~Fabritiis,~G. {TorchMD}-{NET}: Equivariant Transformers for Neural Network based Molecular Potentials. \emph{{arXiv}} \textbf{2022}, {arXiv}:2202.02541, accessed on 2023-07-13\relax
\mciteBstWouldAddEndPuncttrue
\mciteSetBstMidEndSepPunct{\mcitedefaultmidpunct}
{\mcitedefaultendpunct}{\mcitedefaultseppunct}\relax
\EndOfBibitem
\bibitem[Pelaez \latin{et~al.}(2024)Pelaez, Simeon, Galvelis, Mirarchi, Eastman, Doerr, Thölke, Markland, and De~Fabritiis]{pelaez_torchmd-net_2024}
Pelaez,~R.~P.; Simeon,~G.; Galvelis,~R.; Mirarchi,~A.; Eastman,~P.; Doerr,~S.; Thölke,~P.; Markland,~T.~E.; De~Fabritiis,~G. {TorchMD}-Net 2.0: Fast Neural Network Potentials for Molecular Simulations. \emph{J. Chem. Theory Comput.} \textbf{2024}, \emph{20}, 4076--4087\relax
\mciteBstWouldAddEndPuncttrue
\mciteSetBstMidEndSepPunct{\mcitedefaultmidpunct}
{\mcitedefaultendpunct}{\mcitedefaultseppunct}\relax
\EndOfBibitem
\bibitem[Wang \latin{et~al.}(2024)Wang, Wang, Li, He, Li, Wang, Zheng, Shao, and Liu]{wang_enhancing_2024}
Wang,~Y.; Wang,~T.; Li,~S.; He,~X.; Li,~M.; Wang,~Z.; Zheng,~N.; Shao,~B.; Liu,~T.-Y. Enhancing geometric representations for molecules with equivariant vector-scalar interactive message passing. \emph{Nat. Commun.} \textbf{2024}, \emph{15}, 313\relax
\mciteBstWouldAddEndPuncttrue
\mciteSetBstMidEndSepPunct{\mcitedefaultmidpunct}
{\mcitedefaultendpunct}{\mcitedefaultseppunct}\relax
\EndOfBibitem
\bibitem[Batzner \latin{et~al.}(2022)Batzner, Musaelian, Sun, Geiger, Mailoa, Kornbluth, Molinari, Smidt, and Kozinsky]{batzner_e3-equivariant_2022}
Batzner,~S.; Musaelian,~A.; Sun,~L.; Geiger,~M.; Mailoa,~J.~P.; Kornbluth,~M.; Molinari,~N.; Smidt,~T.~E.; Kozinsky,~B. E(3)-equivariant graph neural networks for data-efficient and accurate interatomic potentials. \emph{Nat. Commun.} \textbf{2022}, \emph{13}, 2453\relax
\mciteBstWouldAddEndPuncttrue
\mciteSetBstMidEndSepPunct{\mcitedefaultmidpunct}
{\mcitedefaultendpunct}{\mcitedefaultseppunct}\relax
\EndOfBibitem
\bibitem[Batatia \latin{et~al.}(2022)Batatia, Batzner, Kovács, Musaelian, Simm, Drautz, Ortner, Kozinsky, and Csányi]{batatia_design_2022}
Batatia,~I.; Batzner,~S.; Kovács,~D.~P.; Musaelian,~A.; Simm,~G. N.~C.; Drautz,~R.; Ortner,~C.; Kozinsky,~B.; Csányi,~G. The Design Space of E(3)-Equivariant Atom-Centered Interatomic Potentials. \emph{{arXiv}} \textbf{2022}, {arXiv}:2205.06643, accessed on 2024-04-22\relax
\mciteBstWouldAddEndPuncttrue
\mciteSetBstMidEndSepPunct{\mcitedefaultmidpunct}
{\mcitedefaultendpunct}{\mcitedefaultseppunct}\relax
\EndOfBibitem
\bibitem[Musaelian \latin{et~al.}(2023)Musaelian, Batzner, Johansson, Sun, Owen, Kornbluth, and Kozinsky]{musaelian_learning_2023}
Musaelian,~A.; Batzner,~S.; Johansson,~A.; Sun,~L.; Owen,~C.~J.; Kornbluth,~M.; Kozinsky,~B. Learning local equivariant representations for large-scale atomistic dynamics. \emph{Nat. Commun.} \textbf{2023}, \emph{14}, 579\relax
\mciteBstWouldAddEndPuncttrue
\mciteSetBstMidEndSepPunct{\mcitedefaultmidpunct}
{\mcitedefaultendpunct}{\mcitedefaultseppunct}\relax
\EndOfBibitem
\bibitem[Batatia \latin{et~al.}(2023)Batatia, Kovács, Simm, Ortner, and Csányi]{batatia_mace_2023}
Batatia,~I.; Kovács,~D.~P.; Simm,~G. N.~C.; Ortner,~C.; Csányi,~G. {MACE}: Higher Order Equivariant Message Passing Neural Networks for Fast and Accurate Force Fields. \emph{{arXiv}} \textbf{2023}, {arXiv}:2206.07697, accessed on 2024-04-22\relax
\mciteBstWouldAddEndPuncttrue
\mciteSetBstMidEndSepPunct{\mcitedefaultmidpunct}
{\mcitedefaultendpunct}{\mcitedefaultseppunct}\relax
\EndOfBibitem
\bibitem[Ramakrishnan \latin{et~al.}(2014)Ramakrishnan, Dral, Rupp, and von Lilienfeld]{ramakrishnan_quantum_2014}
Ramakrishnan,~R.; Dral,~P.~O.; Rupp,~M.; von Lilienfeld,~O.~A. Quantum chemistry structures and properties of 134 kilo molecules. \emph{Sci. Data} \textbf{2014}, \emph{1}, 140022\relax
\mciteBstWouldAddEndPuncttrue
\mciteSetBstMidEndSepPunct{\mcitedefaultmidpunct}
{\mcitedefaultendpunct}{\mcitedefaultseppunct}\relax
\EndOfBibitem
\bibitem[Chmiela \latin{et~al.}(2017)Chmiela, Tkatchenko, Sauceda, Poltavsky, Schütt, and Müller]{chmiela_machine_2017}
Chmiela,~S.; Tkatchenko,~A.; Sauceda,~H.~E.; Poltavsky,~I.; Schütt,~K.~T.; Müller,~K.-R. Machine learning of accurate energy-conserving molecular force fields. \emph{Sci. Adv.} \textbf{2017}, \emph{3}, e1603015\relax
\mciteBstWouldAddEndPuncttrue
\mciteSetBstMidEndSepPunct{\mcitedefaultmidpunct}
{\mcitedefaultendpunct}{\mcitedefaultseppunct}\relax
\EndOfBibitem
\bibitem[Smith \latin{et~al.}(2017)Smith, Isayev, and Roitberg]{smith_ani-1_2017}
Smith,~J.~S.; Isayev,~O.; Roitberg,~A.~E. {ANI}-1: an extensible neural network potential with {DFT} accuracy at force field computational cost. \emph{Chem. Sci.} \textbf{2017}, \emph{8}, 3192--3203\relax
\mciteBstWouldAddEndPuncttrue
\mciteSetBstMidEndSepPunct{\mcitedefaultmidpunct}
{\mcitedefaultendpunct}{\mcitedefaultseppunct}\relax
\EndOfBibitem
\bibitem[Smith \latin{et~al.}(2020)Smith, Zubatyuk, Nebgen, Lubbers, Barros, Roitberg, Isayev, and Tretiak]{smith_ani-1ccx_2020}
Smith,~J.~S.; Zubatyuk,~R.; Nebgen,~B.; Lubbers,~N.; Barros,~K.; Roitberg,~A.~E.; Isayev,~O.; Tretiak,~S. The {ANI}-1ccx and {ANI}-1x data sets, coupled-cluster and density functional theory properties for molecules. \emph{Sci. Data} \textbf{2020}, \emph{7}, 134\relax
\mciteBstWouldAddEndPuncttrue
\mciteSetBstMidEndSepPunct{\mcitedefaultmidpunct}
{\mcitedefaultendpunct}{\mcitedefaultseppunct}\relax
\EndOfBibitem
\bibitem[Eastman \latin{et~al.}(2023)Eastman, Behara, Dotson, Galvelis, Herr, Horton, Mao, Chodera, Pritchard, Wang, De~Fabritiis, and Markland]{eastman_spice_2023}
Eastman,~P.; Behara,~P.~K.; Dotson,~D.~L.; Galvelis,~R.; Herr,~J.~E.; Horton,~J.~T.; Mao,~Y.; Chodera,~J.~D.; Pritchard,~B.~P.; Wang,~Y. \latin{et~al.}  {SPICE}, A Dataset of Drug-like Molecules and Peptides for Training Machine Learning Potentials. \emph{Sci. Data} \textbf{2023}, \emph{10}, 11\relax
\mciteBstWouldAddEndPuncttrue
\mciteSetBstMidEndSepPunct{\mcitedefaultmidpunct}
{\mcitedefaultendpunct}{\mcitedefaultseppunct}\relax
\EndOfBibitem
\bibitem[Eastman \latin{et~al.}(2024)Eastman, Pritchard, Chodera, and Markland]{eastman_nutmeg_2024}
Eastman,~P.; Pritchard,~B.~P.; Chodera,~J.~D.; Markland,~T.~E. Nutmeg and {SPICE}: Models and Data for Biomolecular Machine Learning. \emph{J. Chem. Theory Comput.} \textbf{2024}, \relax
\mciteBstWouldAddEndPunctfalse
\mciteSetBstMidEndSepPunct{\mcitedefaultmidpunct}
{}{\mcitedefaultseppunct}\relax
\EndOfBibitem
\bibitem[Sterling and Irwin(2015)Sterling, and Irwin]{sterling_zinc_2015}
Sterling,~T.; Irwin,~J.~J. {ZINC} 15 – Ligand Discovery for Everyone. \emph{J. Chem. Inf. Model.} \textbf{2015}, \emph{55}, 2324--2337\relax
\mciteBstWouldAddEndPuncttrue
\mciteSetBstMidEndSepPunct{\mcitedefaultmidpunct}
{\mcitedefaultendpunct}{\mcitedefaultseppunct}\relax
\EndOfBibitem
\bibitem[Smith \latin{et~al.}()Smith, Nebgen, Lubbers, Isayev, and Roitberg]{smith_less_2018}
Smith,~J.~S.; Nebgen,~B.; Lubbers,~N.; Isayev,~O.; Roitberg,~A.~E. Less is more: Sampling chemical space with active learning. \emph{J. Chem. Phys.} \emph{148}, 241733\relax
\mciteBstWouldAddEndPuncttrue
\mciteSetBstMidEndSepPunct{\mcitedefaultmidpunct}
{\mcitedefaultendpunct}{\mcitedefaultseppunct}\relax
\EndOfBibitem
\bibitem[Devereux \latin{et~al.}(2020)Devereux, Smith, Huddleston, Barros, Zubatyuk, Isayev, and Roitberg]{devereux_extending_2020}
Devereux,~C.; Smith,~J.~S.; Huddleston,~K.~K.; Barros,~K.; Zubatyuk,~R.; Isayev,~O.; Roitberg,~A.~E. Extending the Applicability of the {ANI} Deep Learning Molecular Potential to Sulfur and Halogens. \emph{J. Chem. Theory Comput.} \textbf{2020}, \emph{16}, 4192--4202\relax
\mciteBstWouldAddEndPuncttrue
\mciteSetBstMidEndSepPunct{\mcitedefaultmidpunct}
{\mcitedefaultendpunct}{\mcitedefaultseppunct}\relax
\EndOfBibitem
\bibitem[Chmiela \latin{et~al.}(2023-01-11)Chmiela, Vassilev-Galindo, Unke, Kabylda, Sauceda, Tkatchenko, and Müller]{chmiela_accurate_2023}
Chmiela,~S.; Vassilev-Galindo,~V.; Unke,~O.~T.; Kabylda,~A.; Sauceda,~H.~E.; Tkatchenko,~A.; Müller,~K.-R. Accurate global machine learning force fields for molecules with hundreds of atoms. \emph{Sci. Adv.} \textbf{2023-01-11}, \emph{9}, eadf0873\relax
\mciteBstWouldAddEndPuncttrue
\mciteSetBstMidEndSepPunct{\mcitedefaultmidpunct}
{\mcitedefaultendpunct}{\mcitedefaultseppunct}\relax
\EndOfBibitem
\bibitem[Wu \latin{et~al.}(2018)Wu, Ramsundar, Feinberg, Gomes, Geniesse, Pappu, Leswing, and Pande]{wu_moleculenet_2018}
Wu,~Z.; Ramsundar,~B.; Feinberg,~E.~N.; Gomes,~J.; Geniesse,~C.; Pappu,~A.~S.; Leswing,~K.; Pande,~V. {MoleculeNet}: A Benchmark for Molecular Machine Learning. \emph{{arXiv}} \textbf{2018}, {arXiv}:1703.00564, accessed on 2024-04-24\relax
\mciteBstWouldAddEndPuncttrue
\mciteSetBstMidEndSepPunct{\mcitedefaultmidpunct}
{\mcitedefaultendpunct}{\mcitedefaultseppunct}\relax
\EndOfBibitem
\bibitem[Sorkun \latin{et~al.}(2019)Sorkun, Khetan, and Er]{sorkun_aqsoldb_2019}
Sorkun,~M.~C.; Khetan,~A.; Er,~S. {AqSolDB}, a curated reference set of aqueous solubility and 2D descriptors for a diverse set of compounds. \emph{Sci. Data} \textbf{2019}, \emph{6}, 143\relax
\mciteBstWouldAddEndPuncttrue
\mciteSetBstMidEndSepPunct{\mcitedefaultmidpunct}
{\mcitedefaultendpunct}{\mcitedefaultseppunct}\relax
\EndOfBibitem
\bibitem[Modee \latin{et~al.}(2022)Modee, Laghuvarapu, and Priyakumar]{modee_benchmark_2022}
Modee,~R.; Laghuvarapu,~S.; Priyakumar,~U.~D. Benchmark study on deep neural network potentials for small organic molecules. \emph{J. Comput. Chem.} \textbf{2022}, \emph{43}, 308--318\relax
\mciteBstWouldAddEndPuncttrue
\mciteSetBstMidEndSepPunct{\mcitedefaultmidpunct}
{\mcitedefaultendpunct}{\mcitedefaultseppunct}\relax
\EndOfBibitem
\bibitem[Dwivedi \latin{et~al.}(2022)Dwivedi, Joshi, Luu, Laurent, Bengio, and Bresson]{dwivedi_benchmarking_2022}
Dwivedi,~V.~P.; Joshi,~C.~K.; Luu,~A.~T.; Laurent,~T.; Bengio,~Y.; Bresson,~X. Benchmarking Graph Neural Networks. \emph{{arXiv}} \textbf{2022}, {arXiv}:2003.00982, accessed on 2024-04-24\relax
\mciteBstWouldAddEndPuncttrue
\mciteSetBstMidEndSepPunct{\mcitedefaultmidpunct}
{\mcitedefaultendpunct}{\mcitedefaultseppunct}\relax
\EndOfBibitem
\bibitem[Jumper \latin{et~al.}(2021)Jumper, Evans, Pritzel, Green, Figurnov, Ronneberger, Tunyasuvunakool, Bates, Žídek, Potapenko, Bridgland, Meyer, Kohl, Ballard, Cowie, Romera-Paredes, Nikolov, Jain, Adler, Back, Petersen, Reiman, Clancy, Zielinski, Steinegger, Pacholska, Berghammer, Bodenstein, Silver, Vinyals, Senior, Kavukcuoglu, Kohli, and Hassabis]{jumper_highly_2021}
Jumper,~J.; Evans,~R.; Pritzel,~A.; Green,~T.; Figurnov,~M.; Ronneberger,~O.; Tunyasuvunakool,~K.; Bates,~R.; Žídek,~A.; Potapenko,~A. \latin{et~al.}  Highly accurate protein structure prediction with {AlphaFold}. \emph{Nature} \textbf{2021}, \emph{596}, 583--589\relax
\mciteBstWouldAddEndPuncttrue
\mciteSetBstMidEndSepPunct{\mcitedefaultmidpunct}
{\mcitedefaultendpunct}{\mcitedefaultseppunct}\relax
\EndOfBibitem
\bibitem[Varadi \latin{et~al.}(2022)Varadi, Anyango, Deshpande, Nair, Natassia, Yordanova, Yuan, Stroe, Wood, Laydon, Žídek, Green, Tunyasuvunakool, Petersen, Jumper, Clancy, Green, Vora, Lutfi, Figurnov, Cowie, Hobbs, Kohli, Kleywegt, Birney, Hassabis, and Velankar]{varadi_alphafold_2022}
Varadi,~M.; Anyango,~S.; Deshpande,~M.; Nair,~S.; Natassia,~C.; Yordanova,~G.; Yuan,~D.; Stroe,~O.; Wood,~G.; Laydon,~A. \latin{et~al.}  {AlphaFold} Protein Structure Database: massively expanding the structural coverage of protein-sequence space with high-accuracy models. \emph{Nucleic Acids Res.} \textbf{2022}, \emph{50}, D439--D444\relax
\mciteBstWouldAddEndPuncttrue
\mciteSetBstMidEndSepPunct{\mcitedefaultmidpunct}
{\mcitedefaultendpunct}{\mcitedefaultseppunct}\relax
\EndOfBibitem
\bibitem[{The UniProt Consortium}(2023)]{the_uniprot_consortium_uniprot_2023}
{The UniProt Consortium} {UniProt}: the Universal Protein Knowledgebase in 2023. \emph{Nucleic Acids Res.} \textbf{2023}, \emph{51}, D523--D531\relax
\mciteBstWouldAddEndPuncttrue
\mciteSetBstMidEndSepPunct{\mcitedefaultmidpunct}
{\mcitedefaultendpunct}{\mcitedefaultseppunct}\relax
\EndOfBibitem
\bibitem[Mongan \latin{et~al.}(2007)Mongan, Simmerling, {McCammon}, Case, and Onufriev]{mongan_generalized_2007}
Mongan,~J.; Simmerling,~C.; {McCammon},~J.~A.; Case,~D.~A.; Onufriev,~A. Generalized Born model with a simple, robust molecular volume correction. \emph{J. Chem. Theory Comput.} \textbf{2007}, \emph{3}, 156--169\relax
\mciteBstWouldAddEndPuncttrue
\mciteSetBstMidEndSepPunct{\mcitedefaultmidpunct}
{\mcitedefaultendpunct}{\mcitedefaultseppunct}\relax
\EndOfBibitem
\bibitem[Nguyen \latin{et~al.}(2013)Nguyen, Roe, and Simmerling]{nguyen_improved_2013}
Nguyen,~H.; Roe,~D.~R.; Simmerling,~C. Improved Generalized Born Solvent Model Parameters for Protein Simulations. \emph{J. Chem. Theory Comput.} \textbf{2013}, \emph{9}, 2020--2034\relax
\mciteBstWouldAddEndPuncttrue
\mciteSetBstMidEndSepPunct{\mcitedefaultmidpunct}
{\mcitedefaultendpunct}{\mcitedefaultseppunct}\relax
\EndOfBibitem
\bibitem[Wang \latin{et~al.}(2023)Wang, He, Li, Shao, and Liu]{wang_aimd-chig_2023}
Wang,~T.; He,~X.; Li,~M.; Shao,~B.; Liu,~T.-Y. {AIMD}-Chig: Exploring the conformational space of a 166-atom protein Chignolin with ab initio molecular dynamics. \emph{Sci. Data} \textbf{2023}, \emph{10}, 549\relax
\mciteBstWouldAddEndPuncttrue
\mciteSetBstMidEndSepPunct{\mcitedefaultmidpunct}
{\mcitedefaultendpunct}{\mcitedefaultseppunct}\relax
\EndOfBibitem
\bibitem[Onufriev and Case(2019)Onufriev, and Case]{onufriev_generalized_2019}
Onufriev,~A.~V.; Case,~D.~A. Generalized Born Implicit Solvent Models for Biomolecules. \emph{Annu. Rev. Biophys.} \textbf{2019}, \emph{48}, 275--296\relax
\mciteBstWouldAddEndPuncttrue
\mciteSetBstMidEndSepPunct{\mcitedefaultmidpunct}
{\mcitedefaultendpunct}{\mcitedefaultseppunct}\relax
\EndOfBibitem
\bibitem[Han \latin{et~al.}(2024)Han, Cen, Wu, Li, Kong, Jiao, Yu, Xu, Wu, Wang, Xu, Wei, Liu, Rong, and Huang]{han_survey_2024}
Han,~J.; Cen,~J.; Wu,~L.; Li,~Z.; Kong,~X.; Jiao,~R.; Yu,~Z.; Xu,~T.; Wu,~F.; Wang,~Z. \latin{et~al.}  A Survey of Geometric Graph Neural Networks: Data Structures, Models and Applications. \emph{{arXiv}} \textbf{2024}, {arXiv}:2403.00485, accessed on 2024-08-27\relax
\mciteBstWouldAddEndPuncttrue
\mciteSetBstMidEndSepPunct{\mcitedefaultmidpunct}
{\mcitedefaultendpunct}{\mcitedefaultseppunct}\relax
\EndOfBibitem
\bibitem[Gao \latin{et~al.}(2020)Gao, Ramezanghorbani, Isayev, Smith, and Roitberg]{gao_torchani_2020}
Gao,~X.; Ramezanghorbani,~F.; Isayev,~O.; Smith,~J.~S.; Roitberg,~A.~E. {TorchANI}: A Free and Open Source {PyTorch}-Based Deep Learning Implementation of the {ANI} Neural Network Potentials. \emph{J. Chem. Inf. Model.} \textbf{2020}, \emph{60}, 3408--3415\relax
\mciteBstWouldAddEndPuncttrue
\mciteSetBstMidEndSepPunct{\mcitedefaultmidpunct}
{\mcitedefaultendpunct}{\mcitedefaultseppunct}\relax
\EndOfBibitem
\bibitem[Honda \latin{et~al.}(2008)Honda, Akiba, Kato, Sawada, Sekijima, Ishimura, Ooishi, Watanabe, Odahara, and Harata]{honda_crystal_2008}
Honda,~S.; Akiba,~T.; Kato,~Y.~S.; Sawada,~Y.; Sekijima,~M.; Ishimura,~M.; Ooishi,~A.; Watanabe,~H.; Odahara,~T.; Harata,~K. Crystal Structure of a Ten-Amino Acid Protein. \emph{J. Am. Chem. Soc.} \textbf{2008}, \emph{130}, 15327--15331\relax
\mciteBstWouldAddEndPuncttrue
\mciteSetBstMidEndSepPunct{\mcitedefaultmidpunct}
{\mcitedefaultendpunct}{\mcitedefaultseppunct}\relax
\EndOfBibitem
\bibitem[Kim and Kim(2016)Kim, and Kim]{kim_new_2016}
Kim,~S.; Kim,~H. A new metric of absolute percentage error for intermittent demand forecasts. \emph{Int. J. Forecast.} \textbf{2016}, \emph{32}, 669--679\relax
\mciteBstWouldAddEndPuncttrue
\mciteSetBstMidEndSepPunct{\mcitedefaultmidpunct}
{\mcitedefaultendpunct}{\mcitedefaultseppunct}\relax
\EndOfBibitem
\bibitem[Ingraham \latin{et~al.}(2023)Ingraham, Baranov, Costello, Barber, Wang, Ismail, Frappier, Lord, Ng-Thow-Hing, Van~Vlack, Tie, Xue, Cowles, Leung, Rodrigues, Morales-Perez, Ayoub, Green, Puentes, Oplinger, Panwar, Obermeyer, Root, Beam, Poelwijk, and Grigoryan]{ingraham_illuminating_2023}
Ingraham,~J.~B.; Baranov,~M.; Costello,~Z.; Barber,~K.~W.; Wang,~W.; Ismail,~A.; Frappier,~V.; Lord,~D.~M.; Ng-Thow-Hing,~C.; Van~Vlack,~E.~R. \latin{et~al.}  Illuminating protein space with a programmable generative model. \emph{Nature} \textbf{2023}, \emph{623}, 1070--1078\relax
\mciteBstWouldAddEndPuncttrue
\mciteSetBstMidEndSepPunct{\mcitedefaultmidpunct}
{\mcitedefaultendpunct}{\mcitedefaultseppunct}\relax
\EndOfBibitem
\bibitem[Eastman \latin{et~al.}(2017)Eastman, Swails, Chodera, {McGibbon}, Zhao, Beauchamp, Wang, Simmonett, Harrigan, Stern, Wiewiora, Brooks, and Pande]{eastman_openmm_2017}
Eastman,~P.; Swails,~J.; Chodera,~J.~D.; {McGibbon},~R.~T.; Zhao,~Y.; Beauchamp,~K.~A.; Wang,~L.-P.; Simmonett,~A.~C.; Harrigan,~M.~P.; Stern,~C.~D. \latin{et~al.}  {OpenMM} 7: Rapid development of high performance algorithms for molecular dynamics. \emph{PLOS Comput. Biol.} \textbf{2017}, \emph{13}, e1005659\relax
\mciteBstWouldAddEndPuncttrue
\mciteSetBstMidEndSepPunct{\mcitedefaultmidpunct}
{\mcitedefaultendpunct}{\mcitedefaultseppunct}\relax
\EndOfBibitem
\bibitem[Best \latin{et~al.}(2012)Best, Zhu, Shim, Lopes, Mittal, Feig, and {MacKerell}]{best_optimization_2012}
Best,~R.~B.; Zhu,~X.; Shim,~J.; Lopes,~P. E.~M.; Mittal,~J.; Feig,~M.; {MacKerell},~A. D.~J. Optimization of the Additive {CHARMM} All-Atom Protein Force Field Targeting Improved Sampling of the Backbone $\phi$, $\psi$ and Side-Chain $\chi_1$ and $\chi_2$ Dihedral Angles. \emph{J. Chem. Theory Comput.} \textbf{2012}, \emph{8}, 3257--3273\relax
\mciteBstWouldAddEndPuncttrue
\mciteSetBstMidEndSepPunct{\mcitedefaultmidpunct}
{\mcitedefaultendpunct}{\mcitedefaultseppunct}\relax
\EndOfBibitem
\bibitem[Paszke \latin{et~al.}(2019)Paszke, Gross, Massa, Lerer, Bradbury, Chanan, Killeen, Lin, Gimelshein, Antiga, and et~al.]{paszke_pytorch_2019}
Paszke,~A.; Gross,~S.; Massa,~F.; Lerer,~A.; Bradbury,~J.; Chanan,~G.; Killeen,~T.; Lin,~Z.; Gimelshein,~N.; Antiga,~L. \latin{et~al.}  {PyTorch}: An Imperative Style, High-Performance Deep Learning Library. \emph{{arXiv}} \textbf{2019}, {arXiv}:1912.01703, accessed on 2023-06-12\relax
\mciteBstWouldAddEndPuncttrue
\mciteSetBstMidEndSepPunct{\mcitedefaultmidpunct}
{\mcitedefaultendpunct}{\mcitedefaultseppunct}\relax
\EndOfBibitem
\bibitem[Fey and Lenssen(2019)Fey, and Lenssen]{fey_fast_2019}
Fey,~M.; Lenssen,~J.~E. Fast Graph Representation Learning with {PyTorch} Geometric. \emph{{arXiv}} \textbf{2019}, arXiv:1903.02428, accessed on 2023-06-12\relax
\mciteBstWouldAddEndPuncttrue
\mciteSetBstMidEndSepPunct{\mcitedefaultmidpunct}
{\mcitedefaultendpunct}{\mcitedefaultseppunct}\relax
\EndOfBibitem
\bibitem[Reuther \latin{et~al.}(2018)Reuther, Kepner, Byun, Samsi, Arcand, Bestor, Bergeron, Gadepally, Houle, Hubbell, Jones, Klein, Milechin, Mullen, Prout, Rosa, Yee, and Michaleas]{reuther2018interactive}
Reuther,~A.; Kepner,~J.; Byun,~C.; Samsi,~S.; Arcand,~W.; Bestor,~D.; Bergeron,~B.; Gadepally,~V.; Houle,~M.; Hubbell,~M. \latin{et~al.}  Interactive supercomputing on 40,000 cores for machine learning and data analysis. 2018 IEEE High Performance extreme Computing Conference (HPEC). 2018; p 1–6\relax
\mciteBstWouldAddEndPuncttrue
\mciteSetBstMidEndSepPunct{\mcitedefaultmidpunct}
{\mcitedefaultendpunct}{\mcitedefaultseppunct}\relax
\EndOfBibitem
\bibitem[Loshchilov and Hutter(2019)Loshchilov, and Hutter]{loshchilov_decoupled_2019}
Loshchilov,~I.; Hutter,~F. Decoupled Weight Decay Regularization. \emph{{arXiv}} \textbf{2019}, {arXiv}:1711.05101, accessed on 2024-04-22\relax
\mciteBstWouldAddEndPuncttrue
\mciteSetBstMidEndSepPunct{\mcitedefaultmidpunct}
{\mcitedefaultendpunct}{\mcitedefaultseppunct}\relax
\EndOfBibitem
\bibitem[Robbins and Monro(1951)Robbins, and Monro]{robbins_stochastic_1951}
Robbins,~H.; Monro,~S. A Stochastic Approximation Method. \emph{Ann. Math. Stat} \textbf{1951}, \emph{22}, 400--407\relax
\mciteBstWouldAddEndPuncttrue
\mciteSetBstMidEndSepPunct{\mcitedefaultmidpunct}
{\mcitedefaultendpunct}{\mcitedefaultseppunct}\relax
\EndOfBibitem
\bibitem[Cleveland(1979)]{cleveland_robust_1979}
Cleveland,~W.~S. Robust Locally Weighted Regression and Smoothing Scatterplots. \emph{J. Am. Stat. Assoc.} \textbf{1979}, \emph{74}, 829--836\relax
\mciteBstWouldAddEndPuncttrue
\mciteSetBstMidEndSepPunct{\mcitedefaultmidpunct}
{\mcitedefaultendpunct}{\mcitedefaultseppunct}\relax
\EndOfBibitem
\end{mcitethebibliography}

\end{document}